\documentclass[graphics,twocolumn, usenatbib]{mn2e}

\usepackage{epsfig} 
\usepackage{graphicx} 
\usepackage{color}

\usepackage{deluxetable} 
\usepackage{aas_macros} 
\usepackage{amssymb}

\newcommand{\enzo}{\textsc{Enzo~}}
\newcommand{\cosmos}{\textsc{Cosmos~}}
\newcommand{\darwin}{\textsc{Darwin~}}

\newcommand{\kms}{\mathrm{\, km \, sec^{-1}}}
 
\newcommand{\msol} {\, M_{\odot}}

\def\etal{{\it et al.}~}

\title[Pathways to massive black holes]{Pathways to massive black holes and compact star clusters 
in pre-galactic dark matter haloes with virial temperatures $\gtrsim$ 10000K}

\author[J.A. Regan \etal] 
{John A. Regan\thanks{E-mail:regan@ast.cam.ac.uk} \& Martin G. Haehnelt \\ \\
$^1$ Institute of Astronomy, Madingley Road, Cambridge CB3 0HA \\}

\begin{document}

\maketitle

\begin{abstract}
Large dynamic range numerical simulations of atomic cooling driven
collapse of gas in pre-galactic DM haloes with $T_{\rm vir} \sim 10000 \, \rm{K}$ 
show that the gas loses 90\,\% and more of its angular momentum 
before rotational support sets in.  In a fraction of these
haloes where the metallicity is low and UV radiation suppresses $H_2$
cooling, conditions are thus very favourable for the rapid build-up 
of massive black holes. Depending on the progression of metal enrichment, 
the continued suppression of $H_2$ cooling by external and internal UV
radiation and the ability to trap the entropy produced by the release
of gravitational energy, the gas at the centre of the halo is expected
to  form a supermassive star, a stellar-mass black hole accreting at  
super-Eddington accretion rates or a compact star-cluster undergoing 
collisional run-away of massive stars at its centre. In all three
cases a massive black hole of initially
modest mass finds itself at the center of a rapid inflow of gas 
with inflow rates of $\gtrsim 1 \, M_{\odot} \, \rm{yr^{-1}}$. The massive black
hole will thus grow quickly to a  mass of $10^5$ to $10^6 \, M_{\odot}$ 
until  further inflow is halted either by consumption of gas by star
formation or by  the increasing energy and momentum feedback  from the
growing massive black hole. Conditions  for the formation of massive 
seed black holes in this way are most  favourable in haloes 
with $T_{\rm vir} \sim 15000 \, \rm{K}$  and  $V_{\rm vir} \sim 20 \,
\kms$ with less massive  haloes not allowing collapse of gas 
by atomic cooling and  more massive  haloes being more prone to fragmentation. 
This should imprint a characteristic mass on the mass spectrum of an early 
population of massive black hole seeds in pre-galactic haloes which will 
later grow into the observed population of supermassive black holes in 
galactic bulges. 
\end{abstract}



\begin{table*} \centering
\begin{minipage}{160mm}
\begin{tabular}{ |l |l |l |l | l | l |l } 
\hline \hline \hline
\textbf{\em Sim} & \textbf{\em Boxsize} & \textbf{\em $\bf z_{\rm
init}$} & \textbf{\em $\bf z_{\rm coll}$} & \textbf{\em DM mass} &
\textbf{\em $\bf \Delta$ R} \\  & \textbf{\em [Comoving $\bf h^{-1}$
Mpc]} & & &  \textbf{\em $\bf [M_{\odot}]$}  & \textbf{\em [pc]} \\

\hline  

A & 5.0 & 200.0 & 15.0 & $9.58 \times 10^3$& $9.59 \times 10^{-3}$  \\  
B & 2.0 & 250.0 & 14.0 & $4.42 \times 10^2$& $1.61 \times 10^{-2}$  \\  
C & 2.0 & 250.0 & 15.0 & $4.42 \times 10^2$& $1.61 \times 10^{-2}$  \\ 
D & 10.0 & 175.0 & 15.0 & $5.52 \times 10^4$& $1.04 \times 10^{-2}$  \\  
E & 0.5 & 250.0 & - & $6.90 \times 10^0$ & $1.00 \times 10^{-1}$  \\

\hline  \hline \textbf{\em $\bf M_{\rm tot}$} & \textbf{\em $\bf
R_{200}$ }  & \textbf{\em $\bf V_{200}$ }  & \textbf{\em $\bf T_{\rm
vir}$}  & \textbf{\em $\bf \rho_{\rm max}$} & \textbf{\em $\bf
\lambda$} & \textbf{\em $\bf T_{\rm core}$} \\  \textbf{\em $\bf
[M_{\odot}]$} & \textbf{\em [kpc]} & \textbf{\em [km $\bf sec^{-1}$]}
& \textbf{\em [K]}  & \textbf{\em $\bf [cm^{-3}]$} & & \textbf{\em
[K]} \\

\hline 
$2.64 \times 10^8$  & $1.28 \times 10^3$ & 29.94 & 32276.80 & $1.24 \times 10^9$ & $0.031$ & $6.11 \times 10^3$  \\ 
$5.37 \times 10^7$  & $0.64 \times 10^3$ & 18.99 & 12986.78 & $5.86 \times 10^8$ & $0.026$ & $6.35 \times 10^3$  \\  
$5.15 \times 10^7$  & $0.59 \times 10^3$ & 19.33 & 13455.27 & $7.12 \times 10^8$ & $0.050$ & $6.13 \times 10^3$  \\  
$9.75 \times 10^8$  & $2.18 \times 10^3$ & 51.53 & 95412.04 & $7.64 \times 10^5$ & $0.019$ & $6.45 \times 10^3$  \\  
$9.27 \times 10^5$  & $0.13 \times 10^3$ & 5.52 & 1098.07 & $5.17 \times 10^0$ & $0.026$ & $1.34 \times 10^3$  \\  

\hline  \hline
\end{tabular}
\end{minipage}

\caption{Basic properties of the three simulations (A, B and C)
presented in RH08 along with those of two further simulations (D and E): 
boxsize(comoving $h^{-1}$ Mpc),  starting redshift,
collapse redshift,  DM particle mass  ($M_{\odot}$), spatial resolution ($h^{-1}$ pc), 
total mass of the halo ($M_{\odot}$), the virial radius ($h^{-1}$ pc),  circular
velocity  (km $\mathrm {sec^{-1}}$),  virial temperature (K), maximum baryon density in the
halo ($\rm cm^{-3}$), angular momentum parameter $\lambda$, and  temperature at the core of the halo (K). 
All units are physical units, unless explicitly stated otherwise.}

\label{TableSims}
\end{table*}


\section{Introduction}

\cite{Rees_1978} in his famous ``flow chart'' has mapped out many possible
pathways leading to the formation of a massive black hole.
The discovery of very luminous high redshift quasars by the Sloan Digital Sky Survey
at redshifts greater than 6  is widely seen as an important clue in this respect. 
The presence of super-massive black holes (SMBHs) 
as massive as $3\times 10^{9}$ at these early times 
\cite[]{Haiman_2001, Fan_2001,Fan_2006}  requires  
a rapid and efficient build-up of SMBHs,
challenging models based on Eddington limited growth of stellar mass
black holes  \cite[e.g.][]{Volonteri_2003, Volonteri_2005b,  
Haiman_2006, Lodato_2006, Volonteri_2007b}.   
The collapse of 
gas in pre-galactic haloes with virial temperatures
close to but above the threshold for atomic cooling driven collapse 
is perhaps the most promising route to the fast formation of massive
black holes in the mass range of $10^{4} - 10^{6}  \msol$  which then
can act as seeds for the growth of SMBHs 
\cite[e.g.][]{Haehnelt_1993, Umemura_1993, Eisenstein_1995b, Kauffmann_2000, 
Oh_2002, Bromm_2003,  Koushiappas_2004, 
Begelman_2006,   Lodato_2006, Volonteri_2008, Tanaka_2008}.  \\ 
\indent Early attempts to simulate  the collapse of gas in such haloes  numerically confirmed  
analytical arguments that the gas should not fragment efficiently if 
neither metal nor $H_2$ cooling are important \cite[]{Bromm_2003}. 
Recently the advent of adaptive mesh
refinement (AMR) simulations has enabled numerical simulations of atomic
hydrogen driven collapse with a much improved dynamic range.
\cite{Wise_2007} and \nocite{Regan_2008} Regan \& Haehnelt (2008, RH08) used the AMR code  
\enzo \cite[]{Bryan_1995b, Bryan_1997, Norman_1999, OShea_2004} to
perform numerical simulations  of  the  collapse
of metal-free  gas at the centre of dark matter haloes with virial
temperatures  of $\gtrsim$ 10000 K in a cosmological context.  Both groups
found that the gas collapses isothermally, becomes self-gravitating 
and settles into  a  close to isothermal 
density profile within the DM halo. \cite{Wise_2007} pushed the 
resolution of their simulations to study the evolution of the 
innermost mass shells at the very centre of the halo reaching 
sub-Jupiter mass scales. RH08 instead followed  
the dynamical evolution  of a much larger fraction of the gas in the  
halo until it settled into rotational support. RH08 found that between 0.1\% and  
1\% of the gas at the centre  formed a  massive self-gravitating disc.
The discs are surrounded   by the ongoing isothermal  collapse of the outer mass shells  
which their  simulations were not able to follow due to the prohibitively
 short dynamical timescales in the dense regions of the disc at the centre of the haloes.   The
simulations contain a wealth of information regarding the loss of
angular momentum and mass in-fall rates during the isothermal
collapse in these haloes. We here  discuss the implications for the
expected further dynamical evolution of the gas.\\
\indent In \S \ref{numerics} we will summarize the
main numerical results from RH08, discuss the stability of the
self-gravitating discs which have formed and critically assess the
assumption that $H_2$ and metal cooling may be suppressed in 
(some) pre-galactic haloes with virial temperatures close to but above
the threshold for atomic hydrogen cooling driven collapse. In \S
\ref{dynevolution} we use the dynamical evolution of the  gas settling into rotational 
support in our simulations to  make an educated guess about the
further dynamical evolution of the gas in the mass shells outside of
the discs which are still undergoing collapse at the end of our simulations.  
In \S \ref{implications} we discuss the implications for the formation of massive seed
black holes and for the early growth of supermassive black holes.
In \S \ref{conclusions} we discuss  our results and conclusions.

\section{AMR simulations of atomic hydrogen driven isothermal collapse} \label{numerics}

\subsection{The numerical setup}
The setup of the numerical simulations used here is described in
detail in RH08. We will give here a brief summary of the 
salient properties of the simulations. The simulations 
were performed using the adaptive mesh refinement (AMR)
code \enzo originally   developed  by Greg  Bryan  and
Mike  Norman at  the University of  Illinois (Bryan \& Norman 1995b,
Bryan \& Norman 1997,  Norman \& Bryan 1999, O'Shea et al. 2004).
\nocite{Bryan_1995b} \nocite{OShea_2004} \nocite{Bryan_1997}
\nocite{Norman_1999} We used \enzo in its AMR  mode to simulate 
the collapse of the gas at the centre of DM haloes with virial 
temperatures in the range $13000 - 32000$ K assuming that metal and $H_2$ 
cooling were suppressed.  The three simulations of haloes 
with virial velocities of $19, 20 \, \, \rm{and} \, \, 29 \kms$
described in RH08 will form the basis of
our study here. The haloes were
selected from simulations with box-sizes of $2.0 - 5.0 \, h^{-1} \, \rm{Mpc}$. 
The haloes were resolved with  $\sim 30000$ to $\sim 120000$ 
DM particles and the simulations were run with the maximum refinement
level set to 18 (simulation A) and 16  (simulations B \& C)
corresponding to a 
maximum  spatial resolution of $\sim 0.01$ pc in each
simulation.   Further properties of the simulated haloes are listed in
table \ref{TableSims}. In the table the reader will also 
find the properties of two further simulations with smaller and
bigger virial temperatures/velocities which will be discussed  
in \S \ref{characteristicmasses}.

\begin{figure*}
  
  \includegraphics[width=5.8cm]{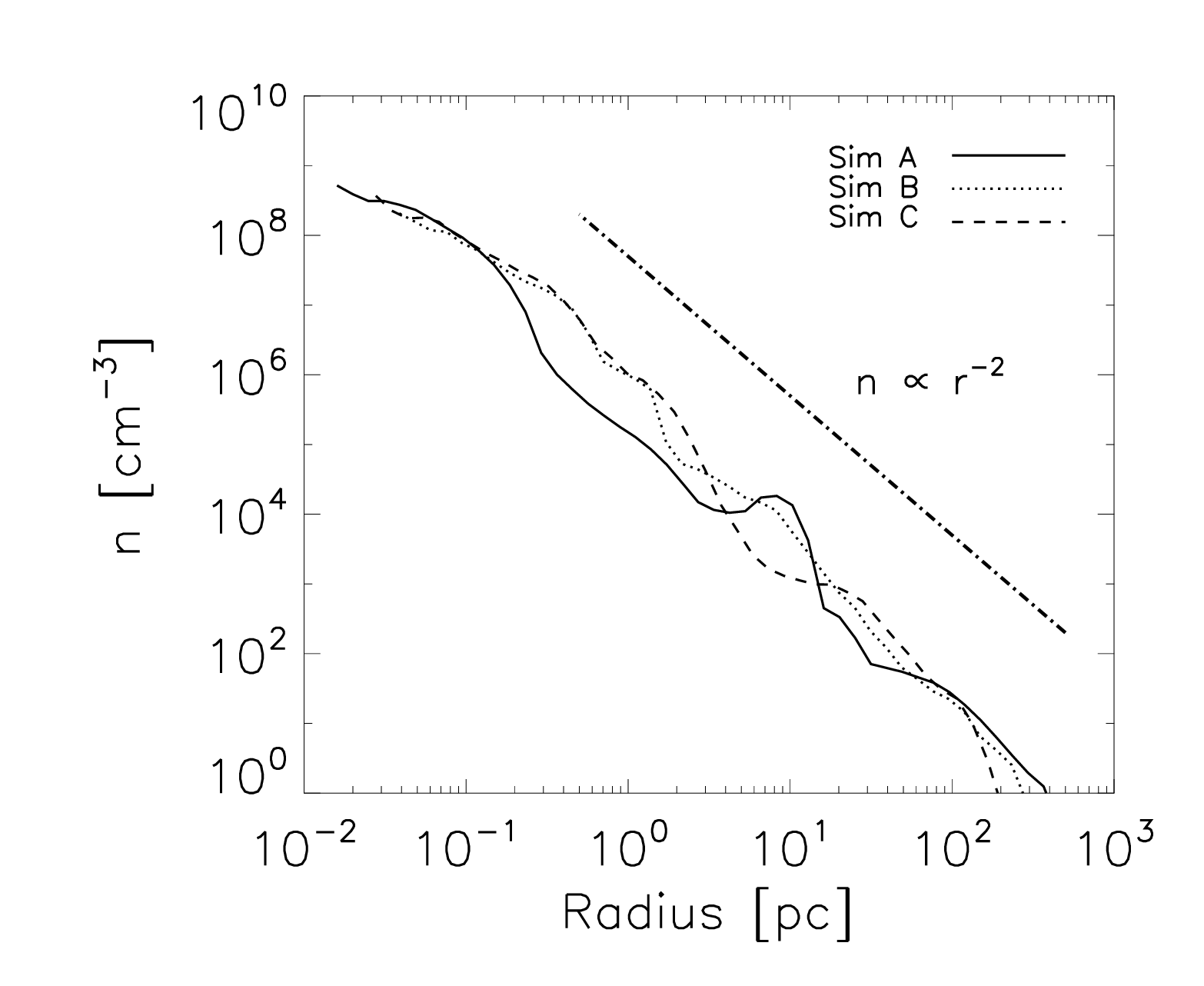}
  \includegraphics[width=5.8cm]{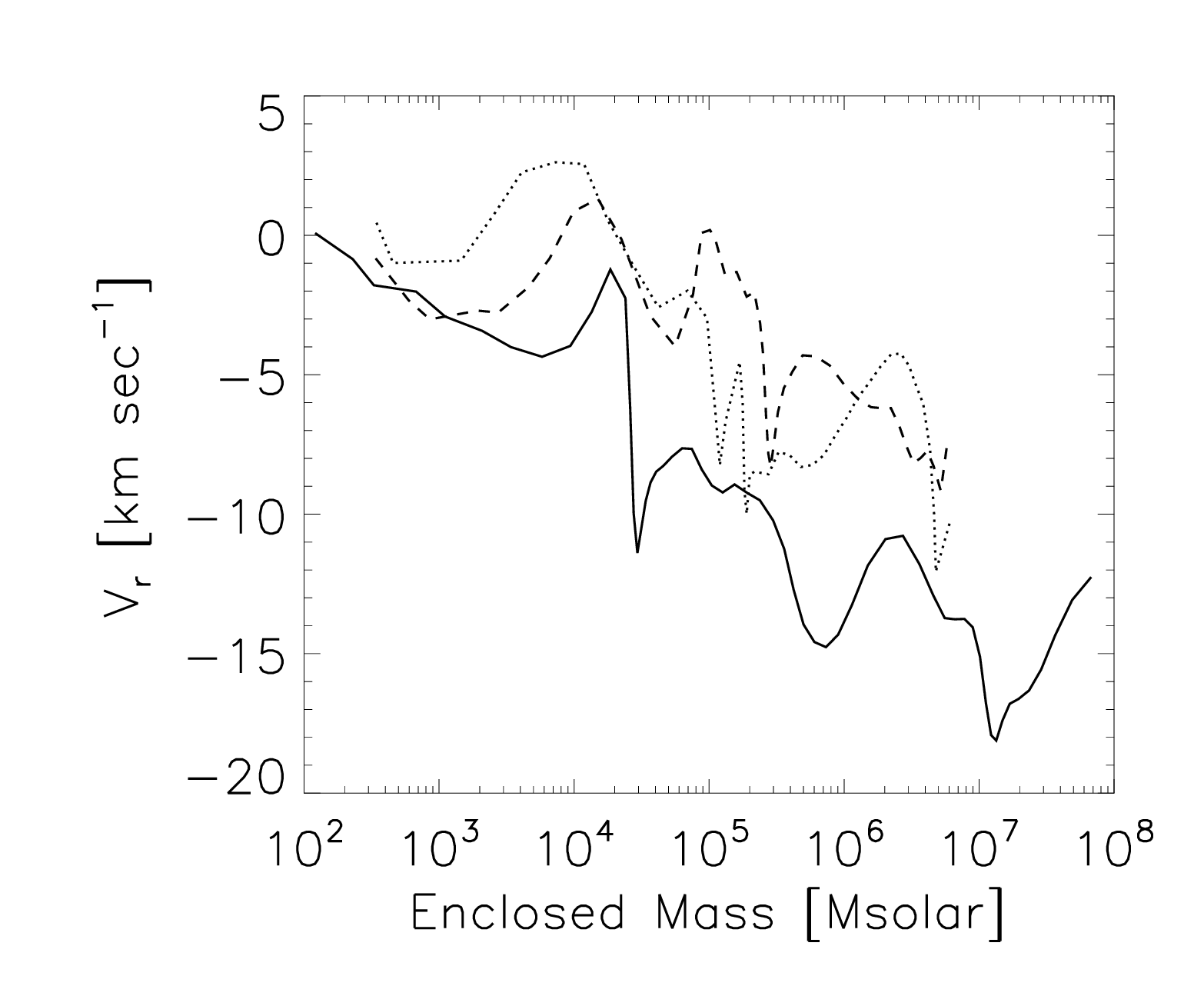}
  \includegraphics[width=5.8cm]{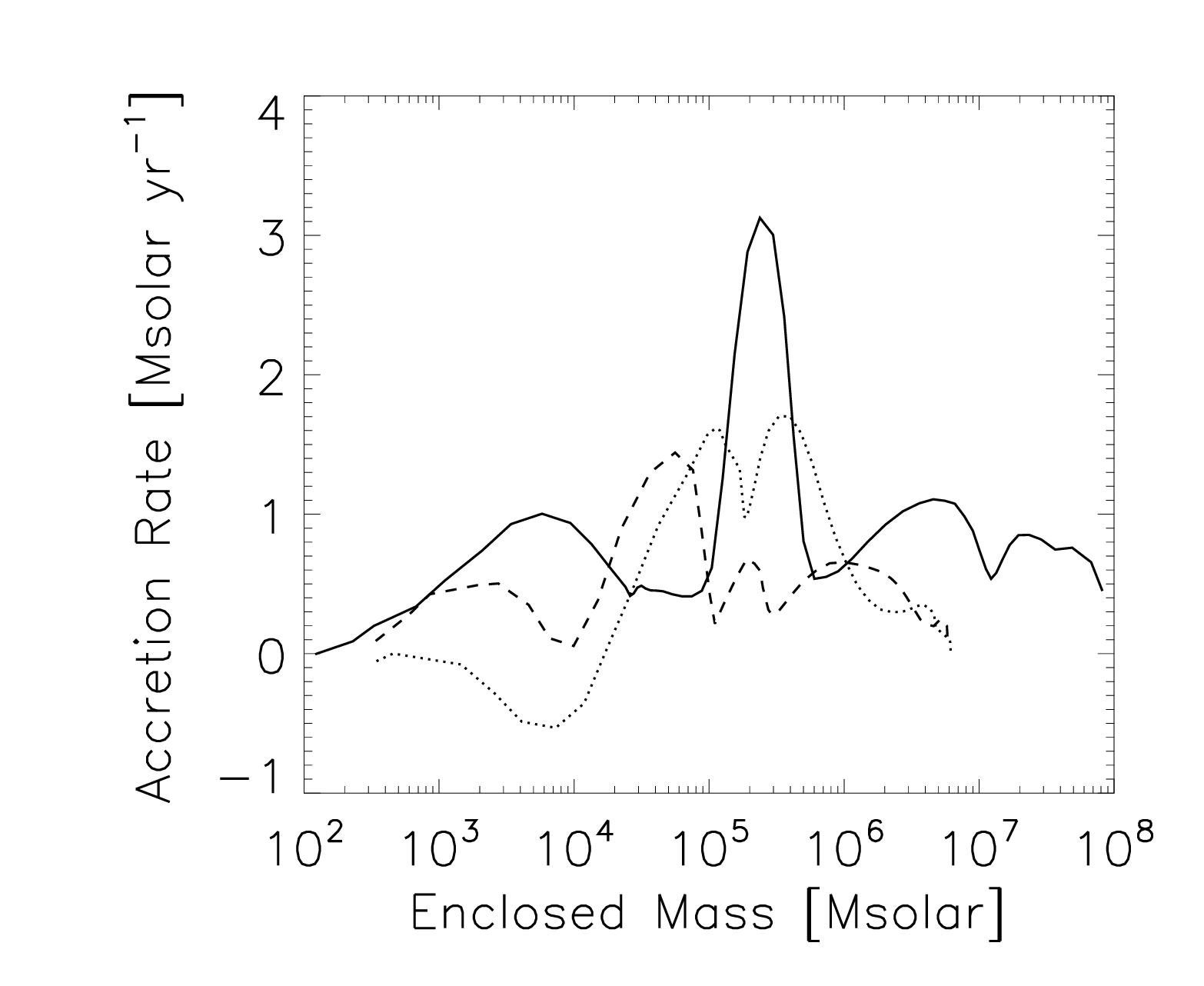}
  \caption[The density, radial velocity and in-fall rate of the
    three simulations.]{\label{infall} {\it Left-hand Panel:} The density
    as a function of radius for simulations A, B \& C. The profile is
    close to  isothermal ($n \propto r^{-2}$) over several decades in radius.
    {\it Middle Panel:} The radial velocity as a function of enclosed gas
    mass. Notice the sharp  drop in the radial velocity at the mass of the
    disc in each simulation.  {\it Right-hand Panel:} The mass in-fall rate for
    each simulation calculated from the density and radial velocity.  }

\end{figure*}
 

\begin{figure*}

  \includegraphics[width=5.8cm]{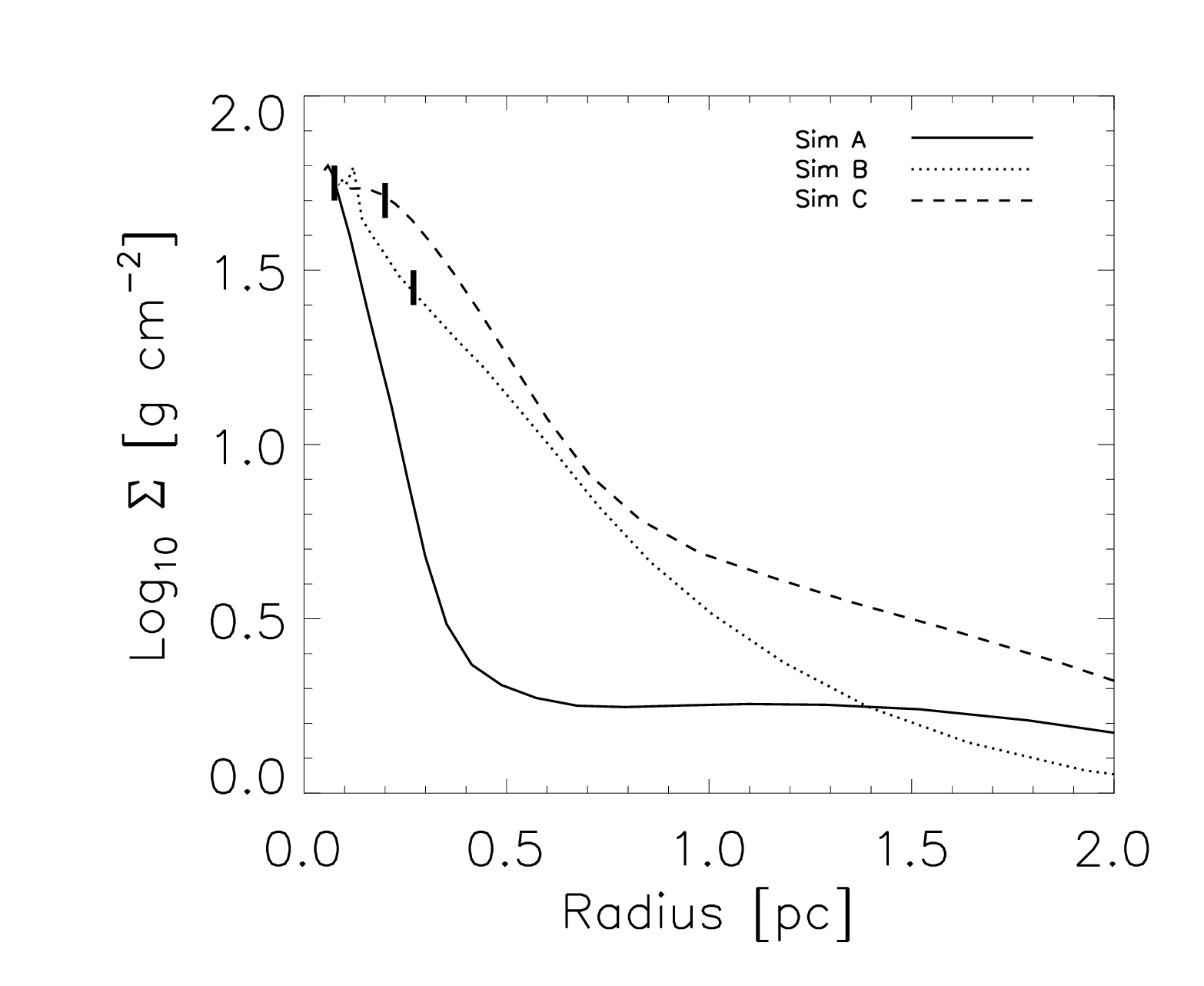}
  \includegraphics[width=5.8cm]{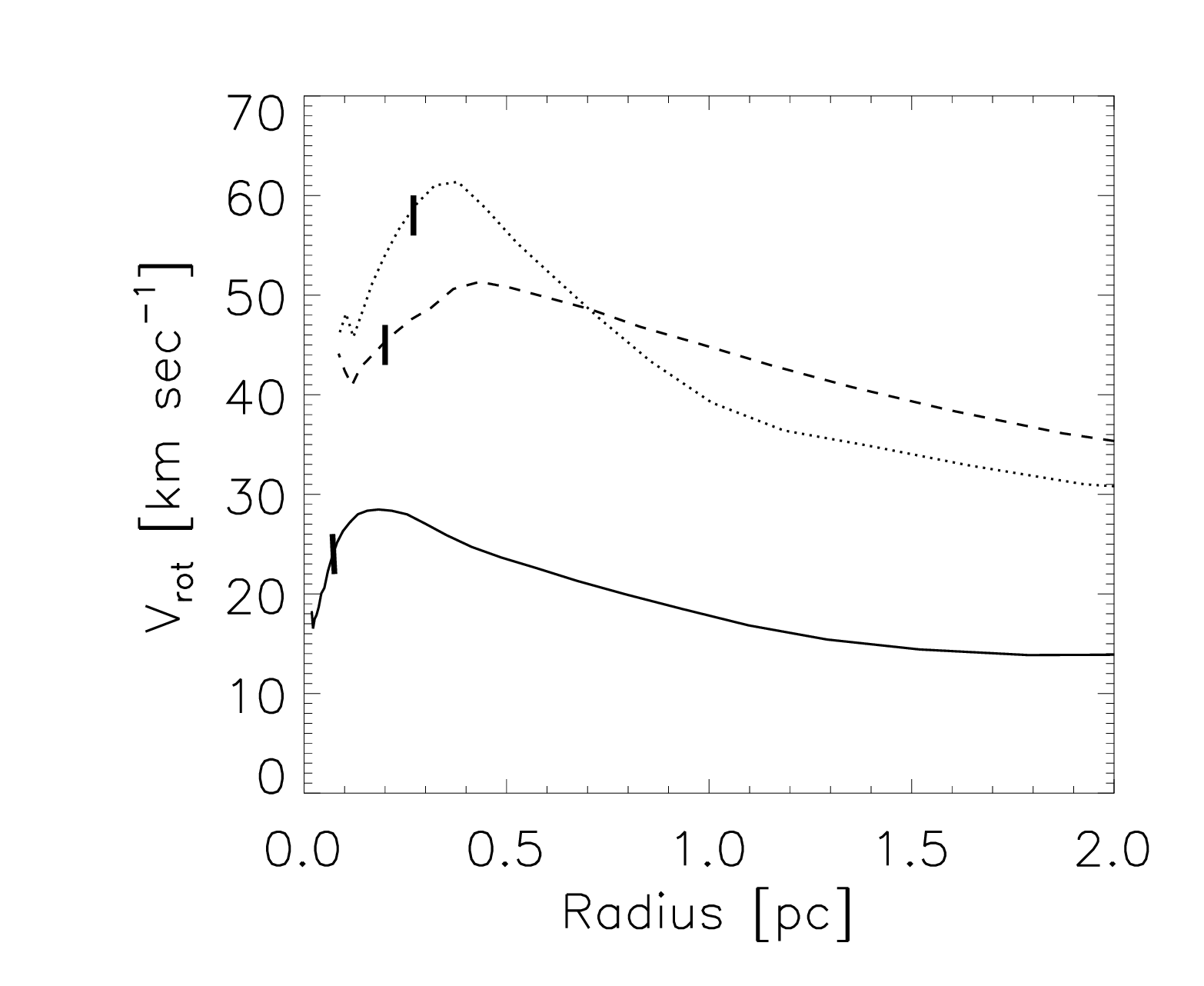}
  \includegraphics[width=5.8cm]{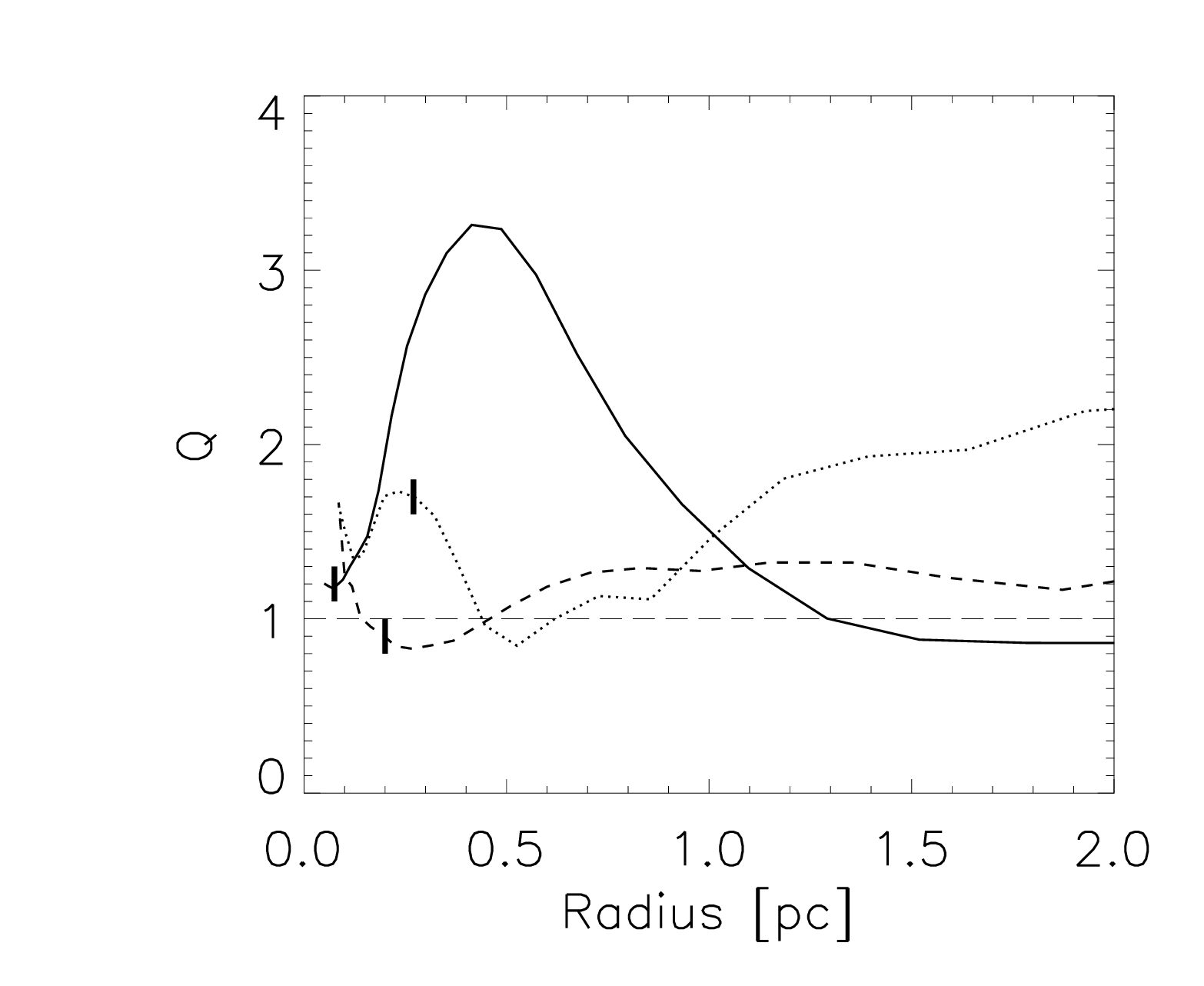}
  \caption[The surface mass density, rotational velocity and the
    Toomre parameter for each disc.]{\label{Toomre} {\it Left-hand Panel:}
    The surface mass density of the disc in simulations A-C. Note the
    exponential  density profile within the disc.  {\it Middle
      Panel:} The rotational velocity of the gas.  {\it Right-hand
      Panel:} The Toomre parameter, $Q$, for the gas in the discs. 
    A value of $Q > 1$ (marked with the long dashed line) 
    indicates that  the disc is  gravitationally stable. 
    The tick marks in each figure indicate the scale radii of
    the discs.  }

\end{figure*}


\subsection{Properties of the Isothermal Collapse}

The gas at the centre of the DM haloes cools to a temperature of $7000 \,-\, 8000$ K due 
to atomic hydrogen cooling and settles  isothermally into a close
to $\rho \propto r^{-2}$  profile as shown in the left panel of Figure \ref{infall}
for simulations A-C. During the collapse, the gas develops supersonic 
turbulent  motions  of order the virial velocity of the DM haloes. The gas
efficiently loses angular momentum and settles towards the centre with  
radial velocities, $V_r$, of $5 - 15 \kms$ about a factor 3-5 smaller than the
virial velocity of the haloes (middle panel of Figure \ref{infall}). The
corresponding mass accretion rates can be estimated as 
\begin{equation}
\dot{M} = 4 \pi r^{2} \rho(r) V_r(r) ,
\end{equation}
where $\rho(r)$ is the density at radius $r$. 
The mass accretion rates range 
from a few tenths to about two solar masses per year (right  panel of Figure \ref{infall}).   \\
\indent The central regions in each halo settle into rotational support and form compact, 
fat, self-gravitating discs with a mass of $ \sim 2\times 10^{4} \msol$
(simulation A) and $\sim 1 \times 10^{5} \msol$ (simulations B and C). Note that
there is a significant drop in the  radial velocity of the gas at the mass
shell marking the outer boundary of the discs.  
The peaks in the mass accretion rate profiles at mass shells outside
of the discs are due to clumps of high density gas within the gas
surrounding  the discs. \\
\indent In Figure \ref{Toomre} we summarize the basic properties of the discs. The discs in
simulations A-C have an exponential surface mass density profile with
scale length of  0.075, 0.2 and 0.27 pc, respectively (left panel).  
In the middle panel of Figure \ref{Toomre} we present the rotation curves of the discs
which show peak rotation velocities in the range from $25 \, -\,  60 \kms$. 
Note that the discs have an elliptical shape and are  somewhat fatter than expected for an ideal
isothermal exponential disc  due to their unrelaxed dynamical
state. 

\section{The further dynamical evolution of the gas} \label{dynevolution}

\subsection{Stability and further dynamical evolution of the discs } \label{timescales}
9

We were able to follow the dynamical evolution of the discs for
several rotation periods and the discs appeared not to fragment during this time. 
The stability of self-gravitating discs is governed by the stabilizing
forces of pressure on small scales and rotation on larges scales. The 
gravitational stability  of discs is normally  characterized  by the
Toomre parameter
\cite[]{Toomre_1964},
\begin{equation}
Q (r) = {c_{\rm{s}} \kappa \over \pi G \Sigma},
\end{equation}  
where $c_{\rm s}$ is the sound speed, $\kappa = \sqrt{2} \, (V_{\rm{rot}}
/ r) \, (1 + \rm{d \, ln \, V_{\rm{rot}}} / \rm{d \, ln \, r})^{1/2} $ \cite[]{Binney_1987, Oh_2002} is the epicyclic frequency, 
$V_{\rm{rot}}$ is the rotational velocity, 
$\Sigma$ is the surface mass density and $r$ is the radius. 
For values
of $Q < 1$ the disc is expected to be gravitationally unstable.  
In the right panel of Figure \ref{Toomre} we show the Toomre parameter
as a function of radius for the three discs in simulations A-C. 
Within a few scale radii  the discs all have $Q \gtrsim 1$.   
The  Toomre   criterion thus predicts that the discs are marginally
stable and do not fragment confirming the visual impression from the
simulations. The further disc evolution should be governed by the viscous time scale. \\
\indent In the framework of modelling the viscous evolution of a self-gravitating
disc as an $\alpha$ disc \cite[]{Shakura_1973}  the viscous time
scale can be approximated as  $t_{\rm visc} = \alpha ^{-1} (H/R)^{-\beta}
t_{\rm dyn} $, where  $t_{\rm dyn}$ is the dynamical time, $(H/R)$ is 
the ratio of disc height to the radius 
of the disc and $\beta = 2$ for a geometrically thin disc \cite[]{Lin_1987}. 
The discs in our simulations are
rather fat  with $(H/R)\sim 0.2-0.3$ and the ``effective'' beta is
probably somewhat smaller than 2.  Initially the $\alpha$ parameter 
should be of order $0.1$  for the marginally stable discs \cite[]{Lin_1990}.  
The viscous timescale on which the disc evolves should  therefore be a few tens of
dynamical times.  Due to the viscous evolution angular momentum will
be transported outwards. The discs will contract, the  rotation
velocity will increase and  gas pressure  support will decrease 
making the discs more violently unstable.  The $\alpha$ parameter will 
thus increase rapidly to order unity and the discs will stay fat due
to the turbulent motions induced by gravitational instabilities. 
Outward angular momentum transport will continue, perhaps, supported by
the bars-in-bars mechanism proposed by  \cite{Shlosman_1989, Shlosman_1990}  (see \cite{Begelman_2006,
Begelman_2008} for a discussion in this context).  The viscous timescale
will then be locked at a few times the ever decreasing 
dynamical time scale.

\begin{figure*}

  \includegraphics[width=8.8cm]{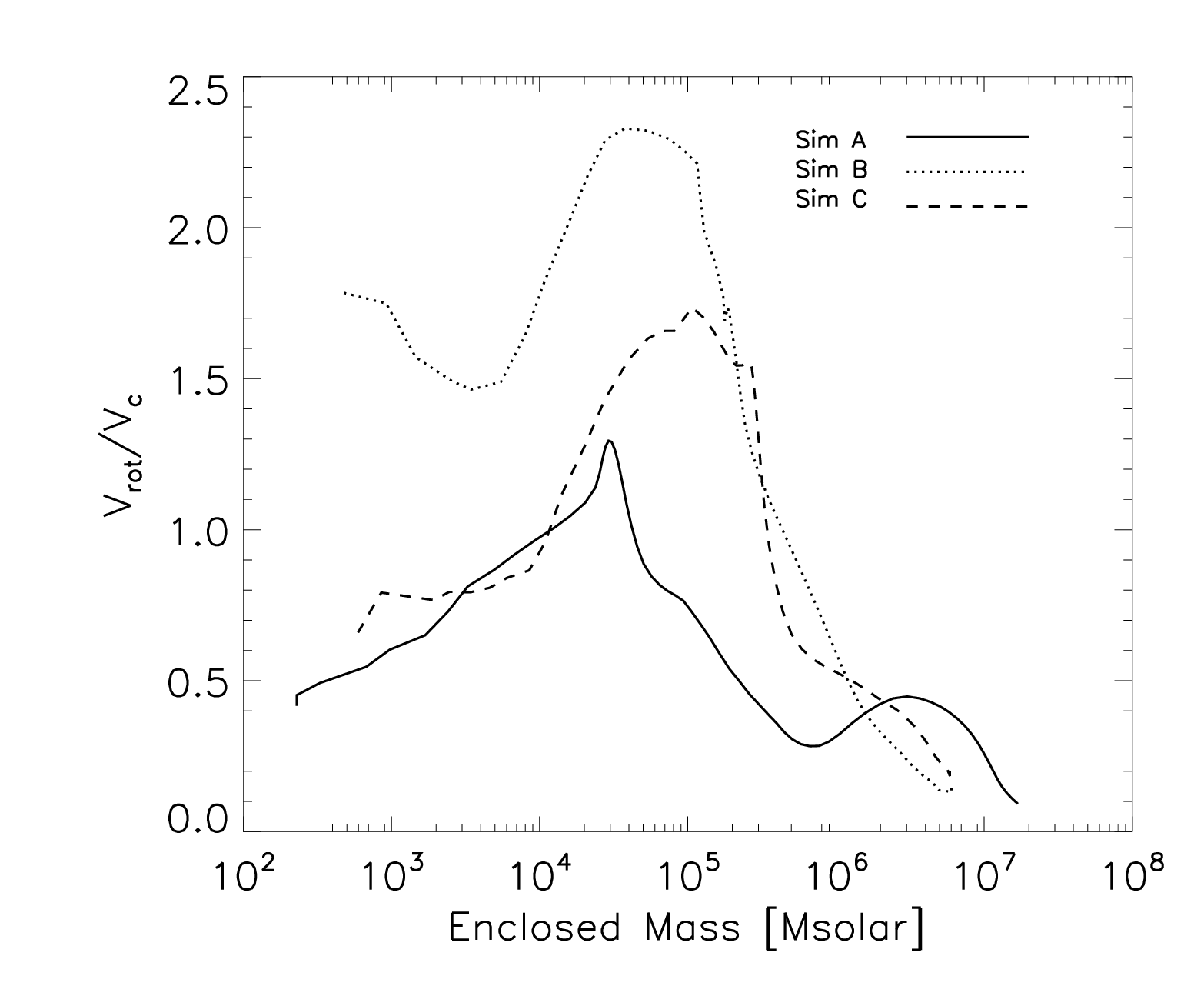}
  \includegraphics[width=8.8cm]{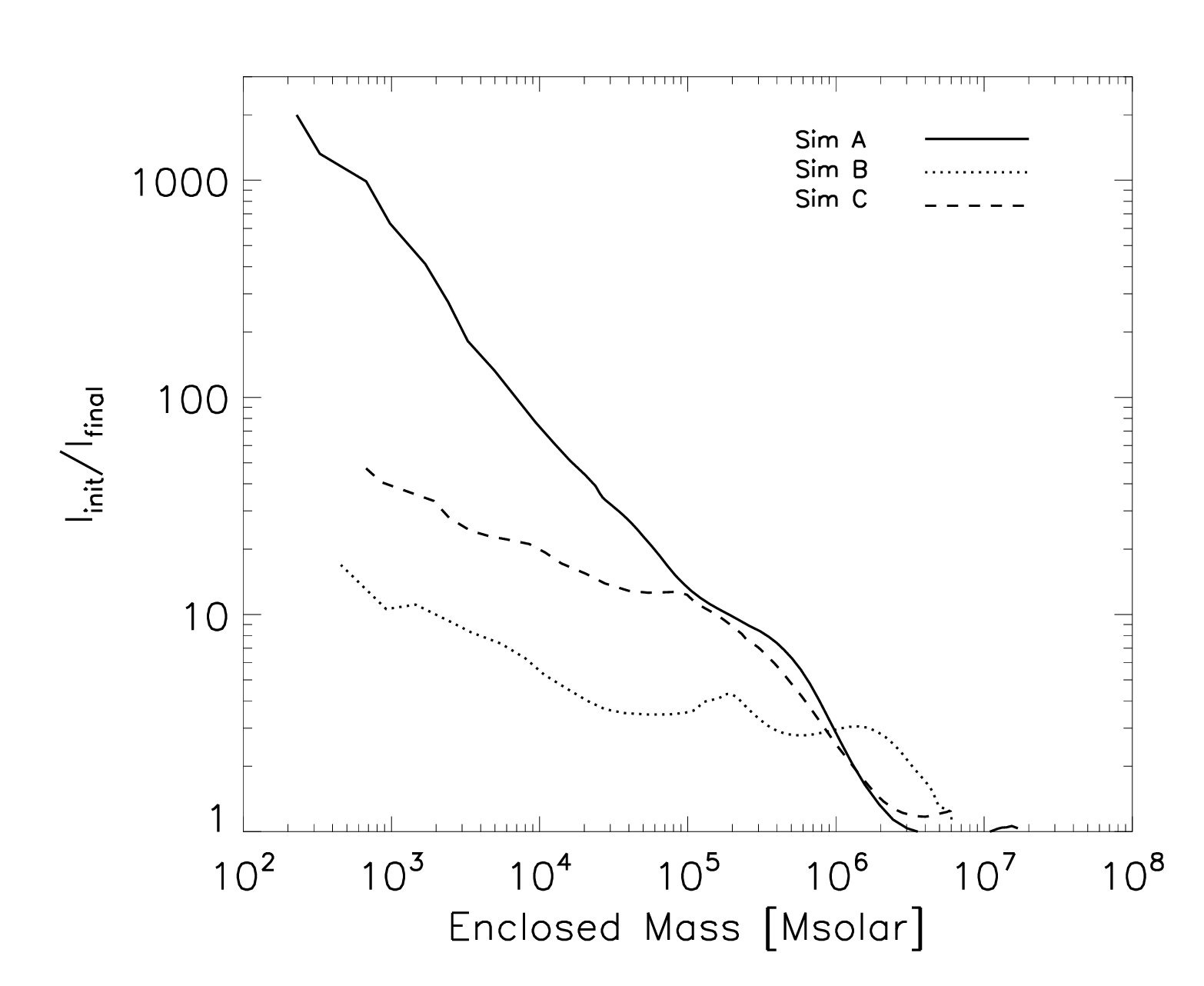}
  \caption[The ratio of the rotational velocity to the circular
    velocity.]{\label{jinitjfinal} {\it Left-hand Panel:} The ratio of the
    rotational velocity, estimated using the inertia tensor (see RH08 for details), 
    to circular velocity,
    $V_c$.  The ratio peaks at the enclosed mass of the disc. A value of
    $\gtrsim 1.0$ indicates  rotational support.  
    {\it Right-hand Panel:} The ratio of the initial angular momentum, $l_{\rm{init}}$, to
    the final angular momentum, $l_{\rm{final}}$.}

\end{figure*}
 

\subsection{ The suppression of metal and $H_2$ cooling  and the role
of star formation} \label{H2cooling}

The most critical assumption in our numerical simulations is that
cooling by atomic hydrogen and
helium is  dominant  which prevents the gas from cooling below $\sim
7000 \, \rm{K}$. The temperature of the gas   close to the virial
temperature of the halo is the reason that no efficient fragmentation
has taken place in the simulations and will presumably also not  occur
during the further collapse of the outer mass shells. The temperature
of the gas is close to the virial temperature of
the discs which is important for their stability.  Cooling by atomic
hydrogen and helium is only dominant when the $H_2$ and metal
abundances  are very small. We will now discuss the plausibility of this
assumption for $H_2$ and metals in turn. \\
\indent  At the densities of the gas in our
simulations the $H_2$ formation timescale is generally shorter than
the dynamical scale. In  the absence  of a strong ultra-violet (UV)
flux $H_2$ should form with a universal abundance of $x_{H_2} \approx
10^{-3}$ \cite[]{Oh_2002}. Efficient dissociation  of $H_2$ is
thus necessary for our assumption, that atomic cooling is dominant,  to
be valid. \\ 
\indent $H_2$ molecules can be dissociated  by
ultra-violet  radiation, either directly by photons with energies
greater than $14.7 \, \rm{eV}$ or as a result of electronic excitation
in the Lyman-Werner (LW)  bands ($ 11.2 \, eV < E < 13.6 \,
eV$). \nocite{Omukai_1999, Glover_2001, Oh_2002, Bromm_2003, Dijkstra_2008} 
Omukai \& Nishi (1999), Glover \& Brand (2001), Oh \& Haiman (2002), Bromm \& Loeb (2003) 
and Dijkstra et al. (2008) have all  discussed in
considerable detail $H_{2}$ formation and dissociation in haloes similar to
the ones we have simulated here.  
UV photons in the LW bands have
substantially longer mean free paths and should thus be more important
for the dissociation of molecular hydrogen. Efficient dissociation of
molecular  hydrogen requires a UV flux  of  $F_{21,LW} \gtrsim 1000$
\cite[]{Bromm_2003},  where $F_{21}$ is in units of $10^{21} \,
\rm{erg \, s^{-1} \, cm^{-2} \, Hz^{-1} \, sr^{-1}}$.  
\cite{Dijkstra_2008} have suggested that there may exist a  small fraction of
haloes close to a neighbouring actively star-forming halo where the
external UV background reaches  this critical flux. However, they 
estimate this fraction to be very small,  a few times
$10^{-7}$.  More important is therefore probably internal UV radiation 
produced within the  haloes. 
Omukai \& Nishi (1999), Glover \& Brand (2001) and Oh \& Haiman (2002)
argue  that as little as one massive star could  in principle be  sufficient to dissociate the 
entire molecular hydrogen  content of our haloes.  The gas in the
discs have rotational velocities of $30-50 \, \kms$ and the gas should thus sit in
sufficiently deep potential wells to withstand the feedback effects 
due to the resulting supernovae.  \\ 
\indent The second assumption we have made in our simulations is
that the halo is virtually metal-free. It is not clear at what metallicity 
metal cooling would become important for our simulations. 
\cite{Omukai_2008} have used one-zone models to study this problem 
and found that  a critical 
metallicity of $Z_{\rm{cr}} \ge 5 \times 10^{-6} \, Z_{\odot}$ leads to a
sufficiently soft equation of state to allow fragmentation via dust 
cooling at densities of $n \sim 10^{10}  \, \rm{cm^{-3}}$, somewhat
higher than the highest densities in our simulations.
In the absence of dust the critical metallicity  rises to  
$Z_{\rm{cr}} \ge 3 \times 10^{-4}\, Z_{\odot}$ above which
fragmentation can occur at substantially lower
densities. It is unclear to what extent  
metals produced by stars  would actually mix
with the gas. If the gas in the haloes were
efficiently enriched with metals fragmentation is obviously
expected to occur. However,  it is then still unclear
how  efficient fragmentation and star formation would be.
Even if the gas in the majority of haloes with $T_{\rm vir} \gtrsim 10000K$ 
fragments and forms an ordinary  star cluster  a small fraction of haloes with
the right conditions to justify our assumption of cooling to be
dominated by atomic cooling may  be  all that is needed to produce the seeds 
for the much smaller number of SMBHs.

\subsection{Further dynamical evolution  of the outer mass shells}

The discs in our simulations sit at the centres of the ongoing
collapse of the gas in the DM haloes.
Unfortunately the dynamical timescales in dense regions of the discs
have become prohibitively short and our simulations  are thus not able
to follow the further dynamical evolution of the outer  mass shells.
As discussed in more detail in RH08 the gas in the outer  parts of the
haloes are neither supported by thermal pressure nor by rotation.  
The gas is instead marginally supported by turbulent pressure and should  thus settle
further  until it either fragments  or reaches  rotational support.
We will now investigate at which radius we should expect  rotational
support  to set in for the gas in the outer mass shells. \\ 
\indent  In the left hand panel of Figure \ref{jinitjfinal} we show an estimate of the
level of rotational support as a  function of enclosed mass at the end
of the simulation. The figure shows the ratio of an estimate of the rotational
velocity  based on the the inertia tensor of the gas  (see RH08 for
details) and the circular velocity, $V_{\rm c}$.  As expected the ratio peaks at
the enclosed mass of the rotationally supported discs for each of the
simulations. Note that in simulation B the ratio  peaks  at an
somewhat  high value of  $\approx 2.4$ due to the complex
dynamical interactions of  two clumps in a multiple clump system. In
the following we will take a ratio of 1.5 as indicative of rotational
support. \\ 
\indent In the right hand panel of Figure
\ref{jinitjfinal} we show the ratio of the initial angular momentum to
the final angular momentum as a function of enclosed mass. We choose
as the initial value of the angular momentum  in each simulation the value at
the time when the maximum refinement level reached is 15 in
simulation A and 12 in simulations B \& C at which time the mass shells
in question are sufficiently resolved.  \\
\indent The loss of angular momentum varies between a factor of 
10 and a factor of 2000 for simulations A to C.  Towards the
outer mass shells  the angular momentum loss by the end of the
numerical simulation becomes progressively  smaller due to the
increasing dynamical time at larger radii. \\
\indent  We first
calculate the radius the outer mass shells would collapse to if
angular momentum  were conserved. As the gas in a given mass shell has
been losing angular momentum  rapidly up to the end of the simulation
this is unlikely to be the case  but should give an upper limit for
the radius at which rotational support is reached. If rotational
support is reached for  $V_{\rm{rot}} (R_{\rm sup}) = 1.5 V_{\rm c }(R_{\rm
sup})$ and angular momentum is conserved the final radius of a mass
shell with enclosed mass, M,  is  related to the radius at the end of
the simulation, $R_{\rm sim}$, as,
\begin{equation} \label{Rsup} R_{\rm{sup}} = \Big[ {V_{\rm{rot}}
(R_{\rm{sim}}) \over V_{\rm{c, gas}}  (R_{\rm{sim}}) \times 1.5}
\Big]^2 \times R_{\rm{sim}} \, .
\end{equation} 
\begin{figure*} 
  \includegraphics[width=8.8cm]{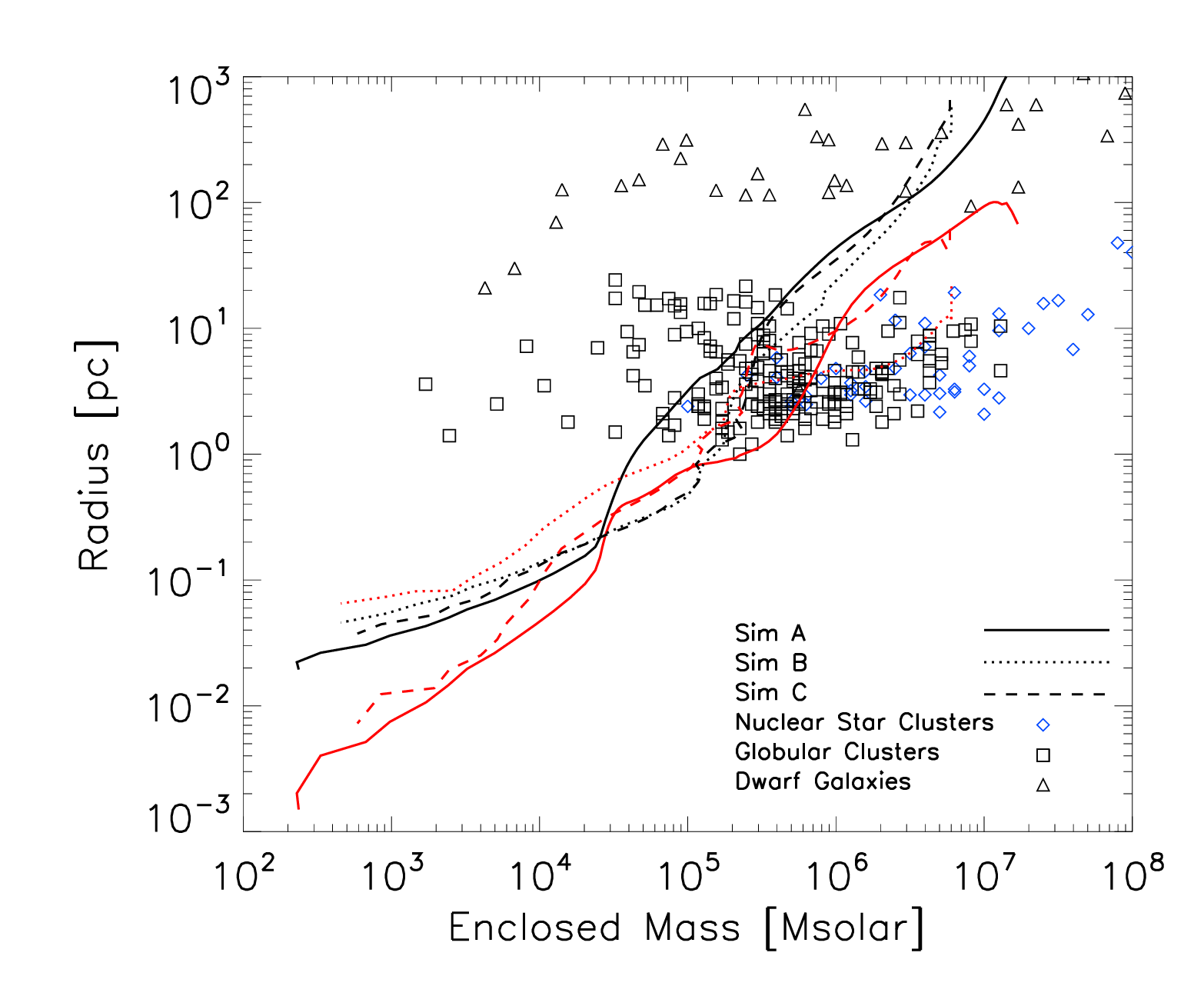}
  \includegraphics[width=8.8cm]{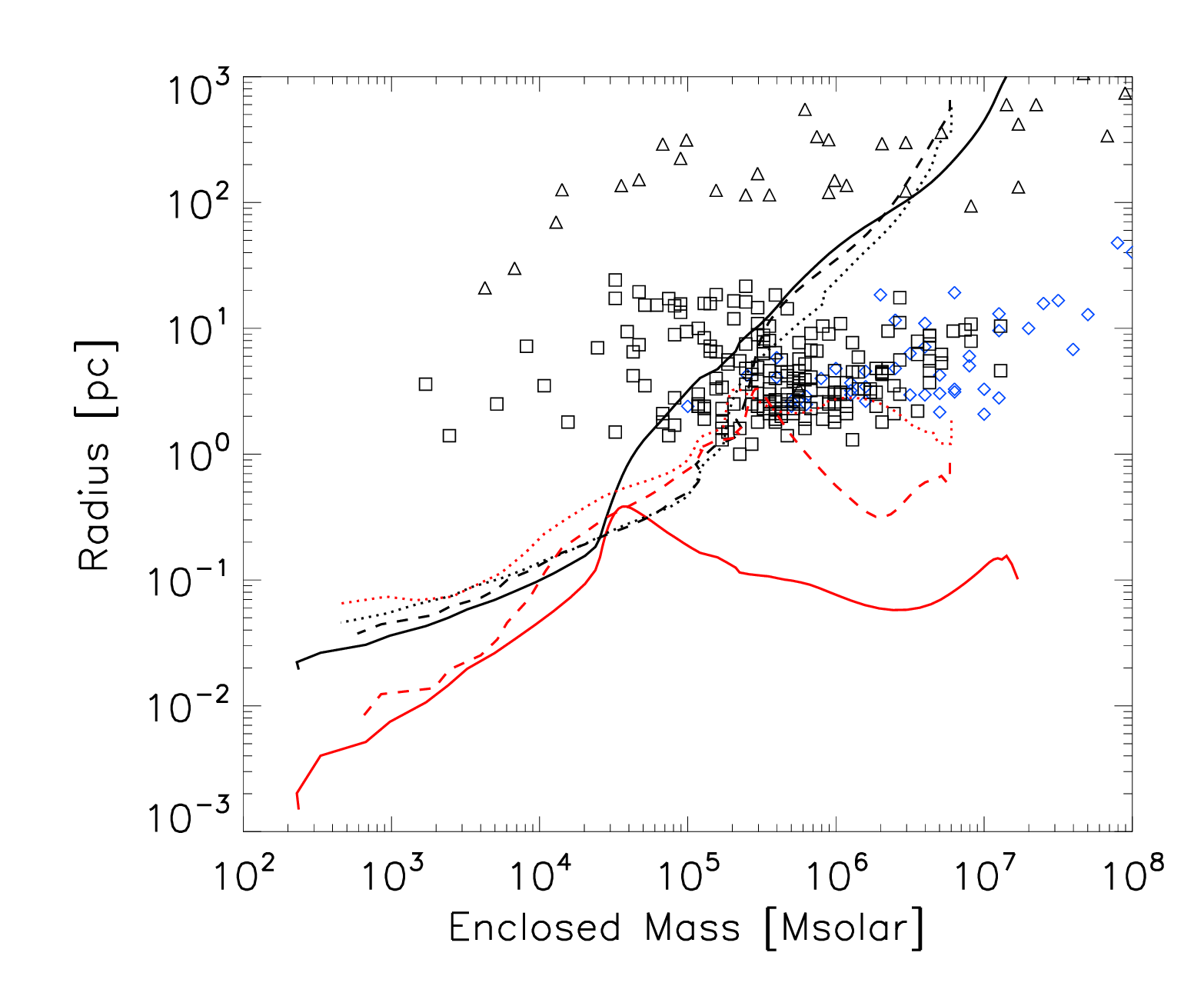}
  \caption[The final radius of the gas.]{\label{Rfinal} {\it Left-hand
Panel:} The black curves show the enclosed gas mass as a function of
radius at the end of simulation A-C. The red curve show the radii to
which the gas will fall if the gas becomes rotationally
supported with $V_{\rm{rot}}(R_{\rm{sup}}) = 1.5 \times V_c(R_{\rm{sup}})$
and angular momentum is conserved.  {\it Right-hand Panel:}
In this case the red curves show the radius to which the gas will
fall if we assume   the gas becomes rotationally supported with
$V_{\rm{rot}}(R_{\rm{sup}}) = 1.5 \times V_c(R_{\rm{sup}})$ and  
the gas loses the same fraction of angular momentum as 
the gas which has settled into rotational support in the disc 
at the end of the simulations. The blue diamonds give masses and radii 
of a sample of observed compact nuclear star clusters as given by 
\cite{Merritt_2008}. The black squares and triangles give masses and radii 
inferred for globular clusters and dwarf galaxies based on data 
from \cite{Belokurov_2007}.  }
\end{figure*}

Note that at the largest radii the gas has not yet become
self-gravitating at the end of the simulation  so $V_{\rm c, gas}
(r_{\rm sim})  = \sqrt{G M_{\rm gas} /r_{\rm sim}}$ is calculated for
the gas mass only.  When rotational support is reached the gas is
always self-gravitating.  In the left hand panel of Figure \ref{Rfinal} we compare the radius as a
function of enclosed mass at the end of the simulation  (black curves)
with  the radius when rotational support is reached as given by equation
\ref{Rsup} (red curves).  Note that for the mass shells within the
discs our estimate for the radius of rotational support can  be larger
than the actual radius at the end of the simulation due to our
assumption  $V_{\rm{rot}} (R_{\rm sup}) = 1.5 V_{\rm c }(R_{\rm{sup}})$
which is exceeded in some parts of the disc.  \\
\indent The gas in the outer mass shells is still rapidly losing
angular momentum at the end of the simulations. We have not fully
understood the angular momentum loss mechanism yet but it appears that
shocks due to the  supersonic turbulent motions lead to the
dissipation of  energy and a redistribution of angular momentum which
is presumably efficiently transported  outwards \cite[]{Wise_2007}.  As we know little
yet about the details of this process we can only speculate about the
further angular momentum loss  of the gas in the outer mass shells. In
the right hand panel of Figure \ref{Rfinal}  we plot the radius
the gas would collapse to if  the gas in all mass shells  loses the same fraction of
angular momentum as that in the outermost mass shell of the disc  which has
formed in the respective simulation.  The radius  where rotational
support sets in  is then given by 
\begin{equation} R_{\rm sup} = \Big[ \beta (M){V_{\rm rot} (R_{\rm sim
}) \over V_{\rm c , gas (R_{\rm sim})} \times 1.5} \Big]^2 \times
R_{\rm sim} ,
\end{equation} 
where $\beta (M)$ is the extra fraction of angular
momentum that must be lost by the outer mass shells so that their
angular momentum loss is the same as that angular momentum lost by the
gas which is rotationally supported at the disc radius. For mass
shells within the disc radius  we set $\beta(M) = 1$.  \\
\indent In order to get a feel
for how compact the configuration of the gas in our simulations is
compared to observed astrophysical objects we also show the typical
masses and radii of observed compact  nuclear star clusters, globular
clusters, and dwarf galaxies \cite[]{Belokurov_2007, Merritt_2008, Seth_2008}.  
We assume a mass-to-light ratio of 5 when converting the 
luminosities  in \cite{Belokurov_2007} to masses.\\
\indent Due to the efficient loss of angular momentum the gas in our
simulations settles into rotational support  in a  substantially more compact 
configuration than the stellar component of nearby dwarf galaxies
 which are hosted by DM haloes with virial
velocities similar to the ones we have simulated here, 
albeit obviously at $z=0$ instead of $z=15$. At the same mass the radii are 
typically one to three orders of magnitude smaller. Observed globular cluster 
and nuclear star cluster are more compact and are generally hosted by galaxies/DM haloes
much more massive than the DM haloes we are considering here
making a comparison less straight forward. 
The inner mass shells ($ \lesssim 10^{5} \, M_{\odot}$) of the gas in our DM haloes 
settle into a configuration which is  also substantially (by 
an order of magnitude in radius) more compact   than globular 
clusters of the same mass. The
configuration of the gas in the outer mass shells is expected to
settle into rotational support at radii which are similar or smaller than
those of compact nuclear star clusters depending on how much angular momentum loss will occur. 
Note, however,  that the observed clusters may have expanded due to the rapid removal 
of gas \cite[]{Bastian_2006}.

\begin{figure*}   
  \includegraphics[height=19cm, width=18cm]{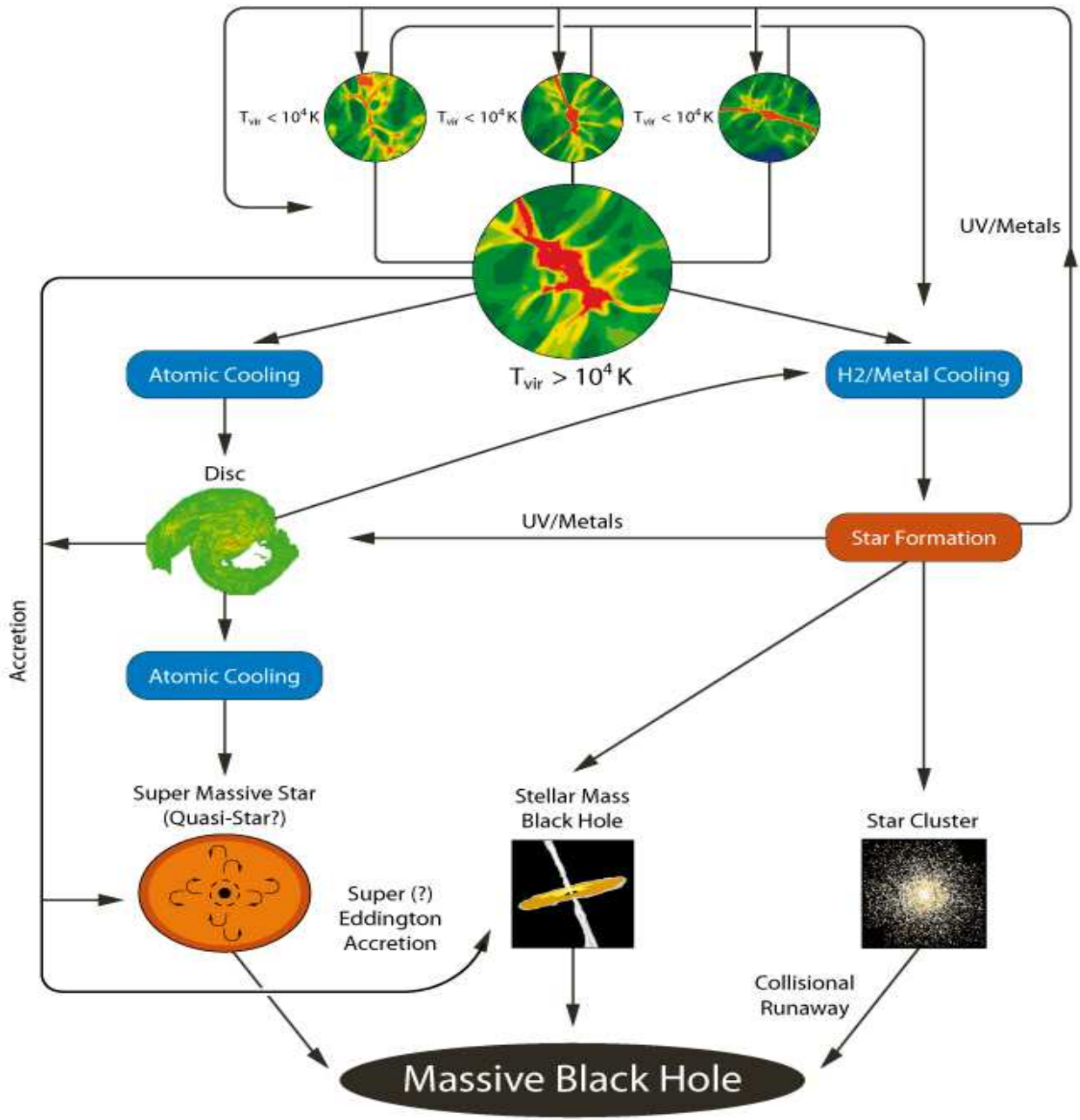}
  
  \caption[Pathways towards a massive black hole]{\label{cartoon}
    Summary of  the possible pathways to massive black
    holes via a stellar seed black hole, a quasi star or via 
    a nuclear star cluster in DM haloes with $T_{\rm vir} \gtrsim 10000 \, \rm{K}.^{1}$ 
   
  }
\end{figure*}
     

\begin{figure*}    
  \includegraphics[width=8.8cm]{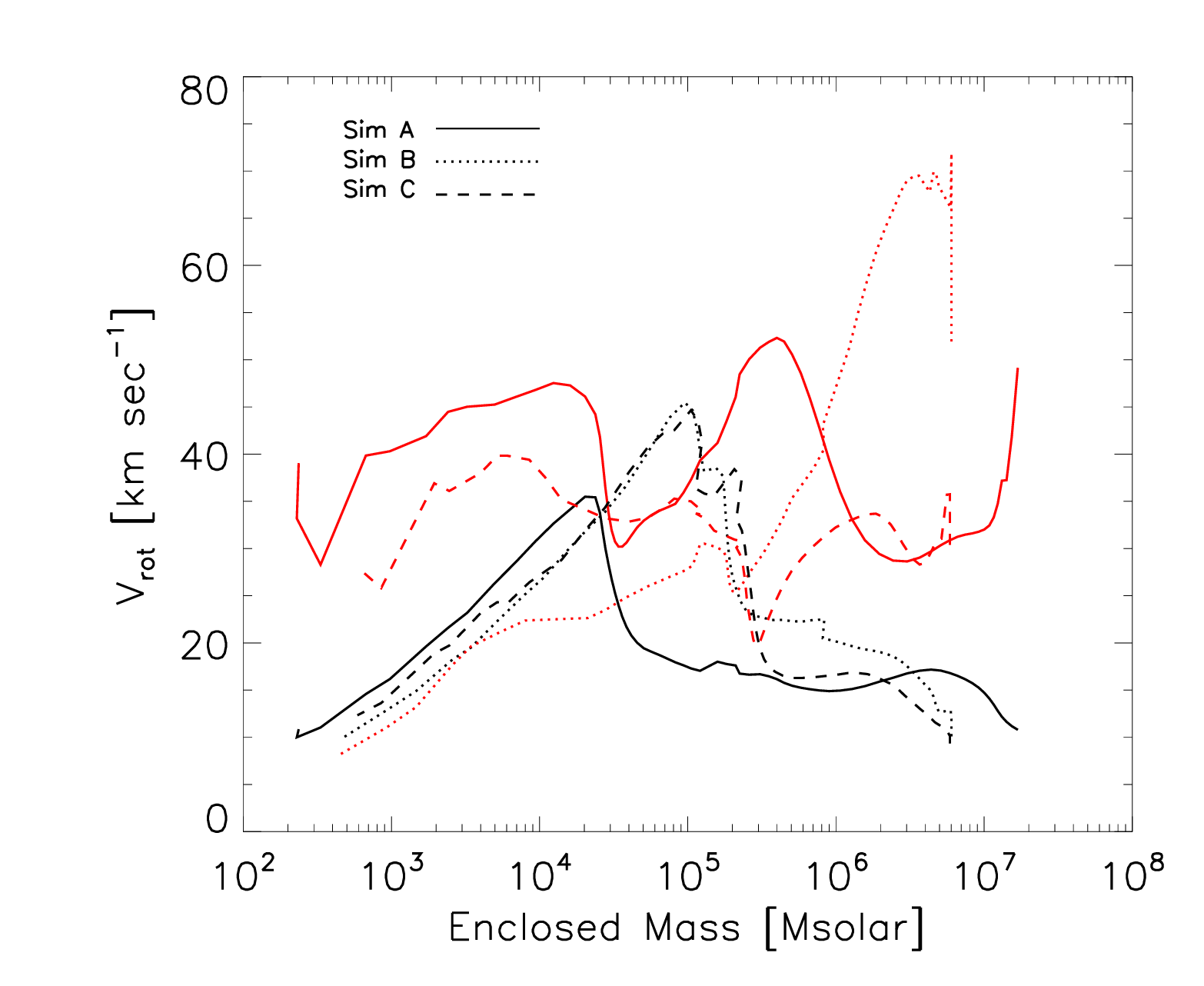}
  \includegraphics[width=8.8cm]{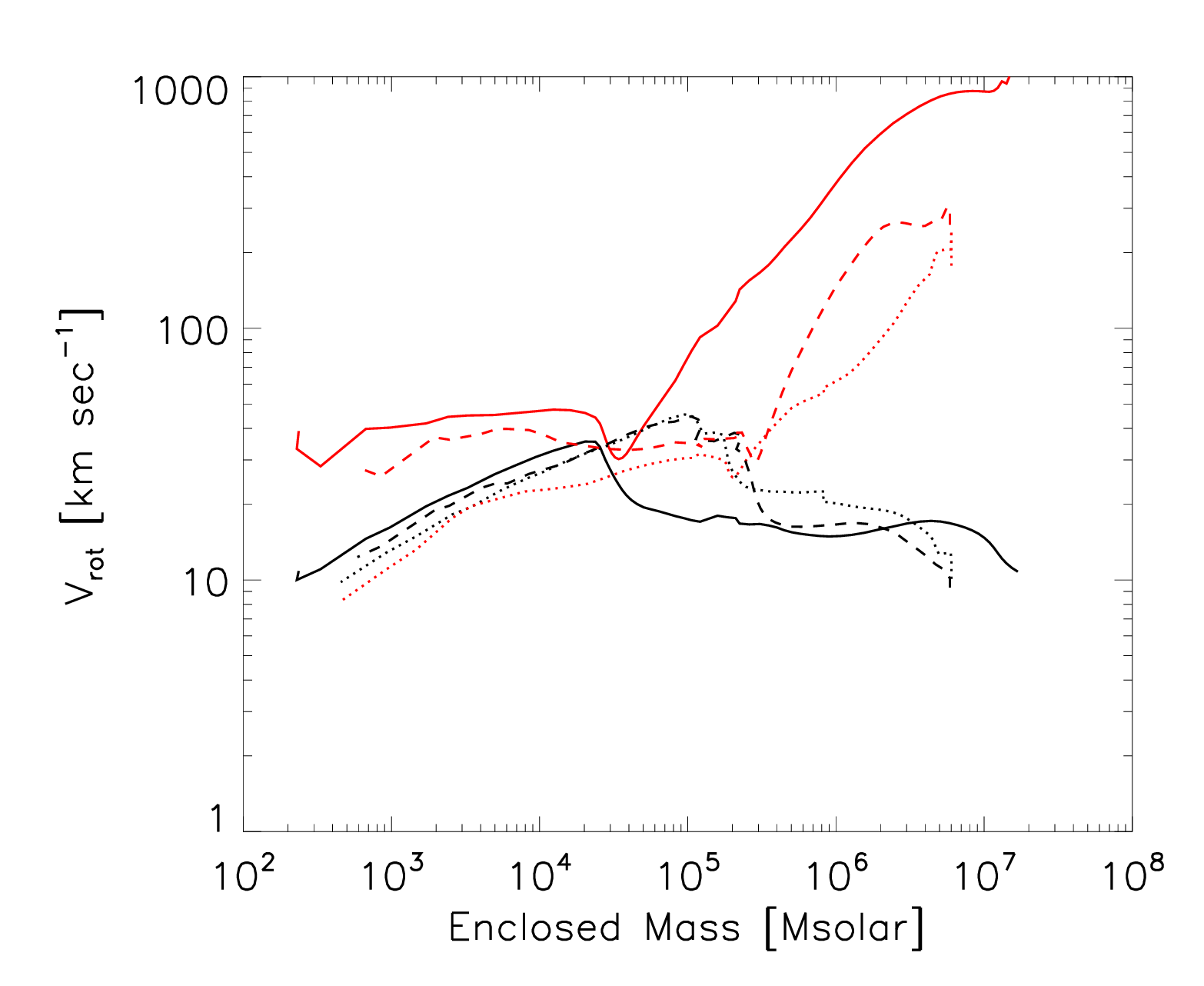}
  \caption[The final rotational velocity of the gas.]{\label{Vcfinal}
    {\it Left-hand Panel:} The black curves show the rotational velocity as a
    function of enclosed mass at the end of simulation A-C. The red curves
    show the rotational velocity  the gas attains if it settles into 
    rotational  support with $V_{\rm{rot}}(R_{\rm{final}}) = 1.5 \times V_c(R_{\rm{final}})$ and
    angular momentum is conserved. The values of $R_{\rm{final}}$
    in Figure \ref{Rfinal} are used.  {\it Right-hand Panel:} In this case
    the red curves show the rotational velocity of the gas attains if we
    assumes the gas reaches rotational support with $V_{\rm{rot}}(R_{\rm{final}}) = 1.5 \times
    V_c(R_{\rm{final}})$ and loses the same fraction of angular momentum as 
    the gas which has settled into rotational support in the disc 
    at the end of the simulations.  The
    values of $R_{\rm{final}}$ in the right hand panel of Figure
    \ref{Rfinal} are  used in this case. }
\end{figure*}
     

\section{Implications for the formation of massive seed black holes} \label{implications}

\subsection{Pathways  to a massive black hole} \label{pathways}

Various mechanisms by which a massive black hole may form 
have been laid out comprehensively by  Rees 
\cite[]{Rees_1978,Rees_1984}. Since then 
considerable effort has gone into researching the paths in this flow
chart but the actual formation process of SMBH nevertheless still
eludes us. Figure \ref{cartoon} summarizes the paths to a massive 
black hole concentrating on the possibilities that pre-galactic DM haloes with 
$T_{\rm vir} \gtrsim 10000 \, \rm{K} $ offer.  We thereby use the results of our 
numerical simulations as a starting point. In the now well established
hierarchical paradigm for galaxy formation  \cite[]{White_1978} 
DM haloes with $T_{\rm vir} \gtrsim 10000 \, \rm{K}$ build-up through the merging 
of several DM haloes with smaller virial temperatures. In these less massive
haloes gas can only collapse and form stars, if the gas cools due to  
$H_2$  and/or metals. Especially for the first generation of these
haloes it is thus very uncertain, how efficiently they have formed stars 
\cite[e.g.][]{Greif_2008, Norman_2008, Whalen_2008}. As discussed in \S \ref{H2cooling}
 this also introduces   considerable uncertainty for the cooling processes  and star formation 
efficiency in the more massive haloes with  $T_{\rm vir} \gtrsim 10000K$.  
If fragmentation and star formation is efficient early on, an ordinary
star cluster/dwarf galaxy may form. If cooling on the other hand is dominated by atomic
cooling the gas is expected instead to settle into a rotationally supported, 
fat, self-gravitating disc \cite[]{Mo_1998, Oh_2002}. The further fate of this disc depends
crucially on whether atomic cooling remains the dominant cooling process 
during the further evolution of the disc. If this is the case the gas
will not efficiently fragment and will continue to contract on a
timescale controlled by the rate at which the gas loses angular
momentum. As the gas will be gravitationally unstable this is expected to
occur on a timescale which is longer by a factor of a few  than the dynamical 
timescale. The further fate then depends on how efficiently the entropy
produced by the release of gravitational energy is trapped. 
\footnotetext[1]{The four pictures of simulated haloes at the top of the figure and the 
    disc show simulations described  in RH08, the sketch of the
    quasi-star is based on that in \cite{Begelman_2008} and the
    picture of the star cluster shows  the 
    globular cluster M80 (source: http://hubblesite.org/newscenter/archive/1999/26/image/a).}
One possibility is that the gas settles  into an high entropy super-massive star
\cite[e.g.][]{Hoyle_1963, Hoyle_1963b, Chandrasekhar_1964, Chandrasekhar_1964b, ZelDovich_1966, ZelDovich_1970, Shapiro_1979, Shapiro_1983} 
which may take the form of a quasi-star
as recently suggested by \cite{Begelman_2006}. If the released
gravitational energy is instead transported away in a disc-jet 
configuration then  rapid accretion onto a stellar mass seed black 
hole would be the expected outcome. Finally if either 
metal or $H_2$ cooling become efficient in the self-gravitating disc
and fragmentation is efficient a very compact star cluster should
form \cite[e.g.][]{Omukai_2008, Bernadetta_2008} . In the following we will discuss these
different possibilities in more detail.

\begin{figure*}

  \includegraphics[width=8.8cm]{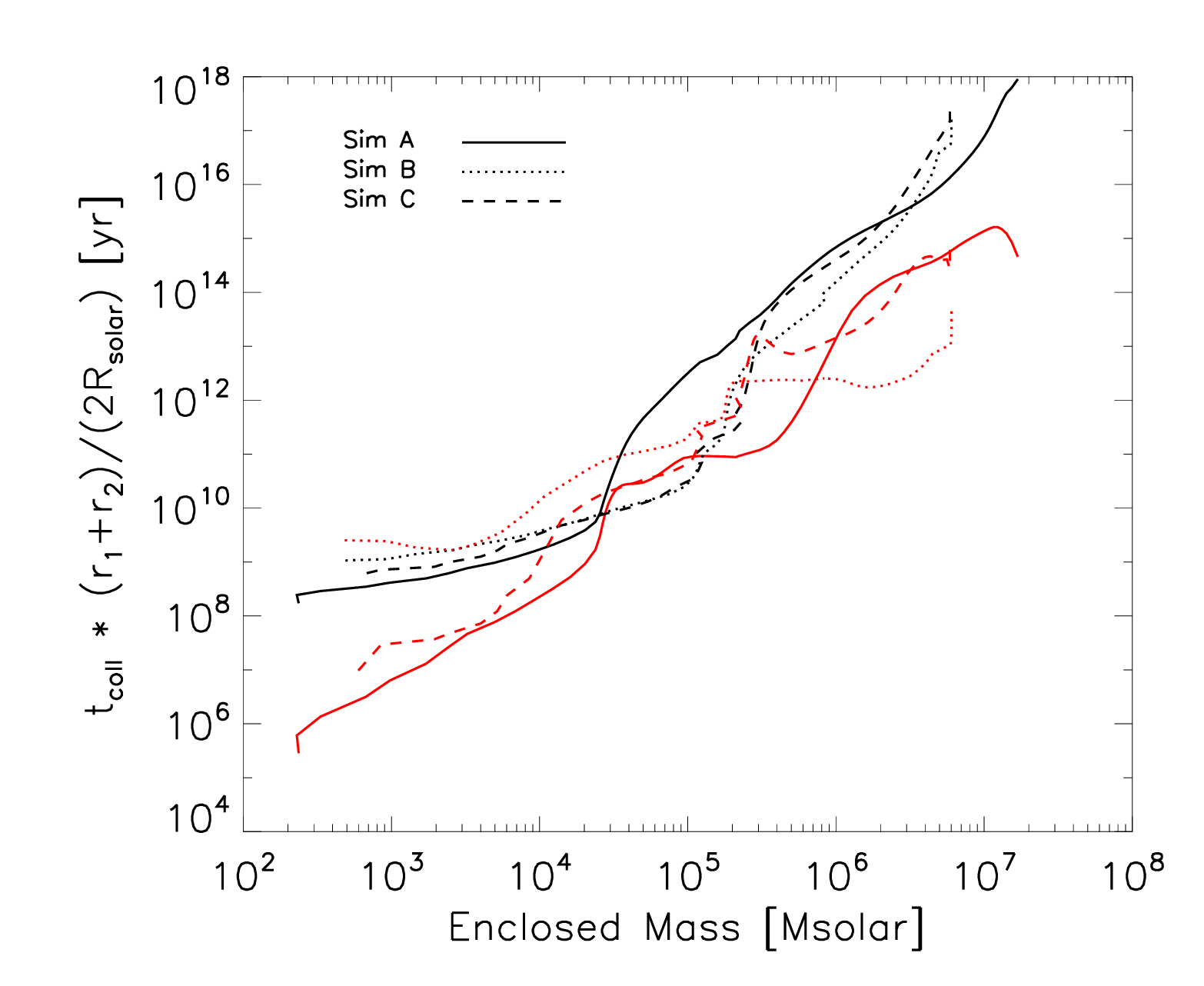}
  \includegraphics[width=8.8cm]{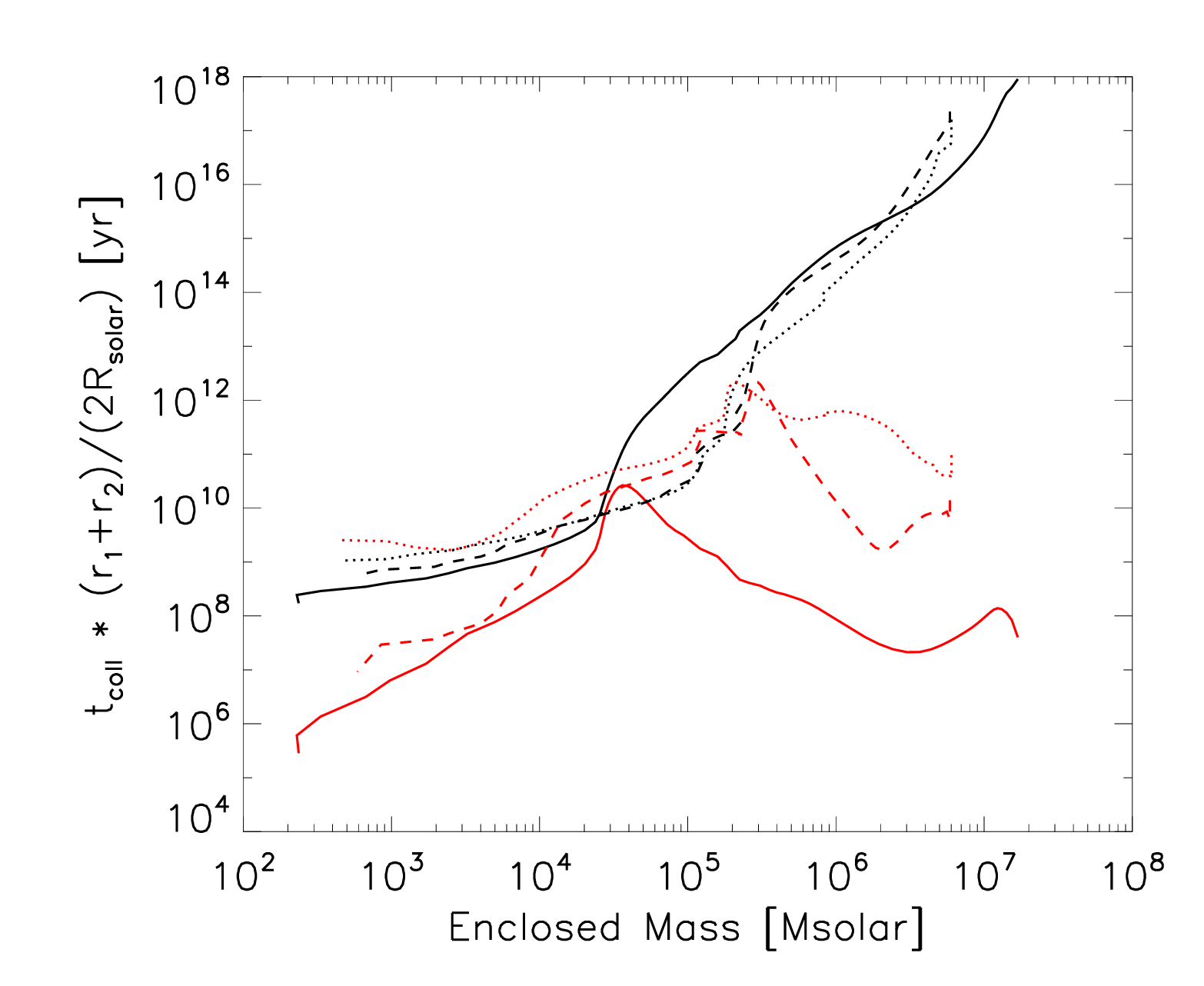}
  \includegraphics[width=8.8cm]{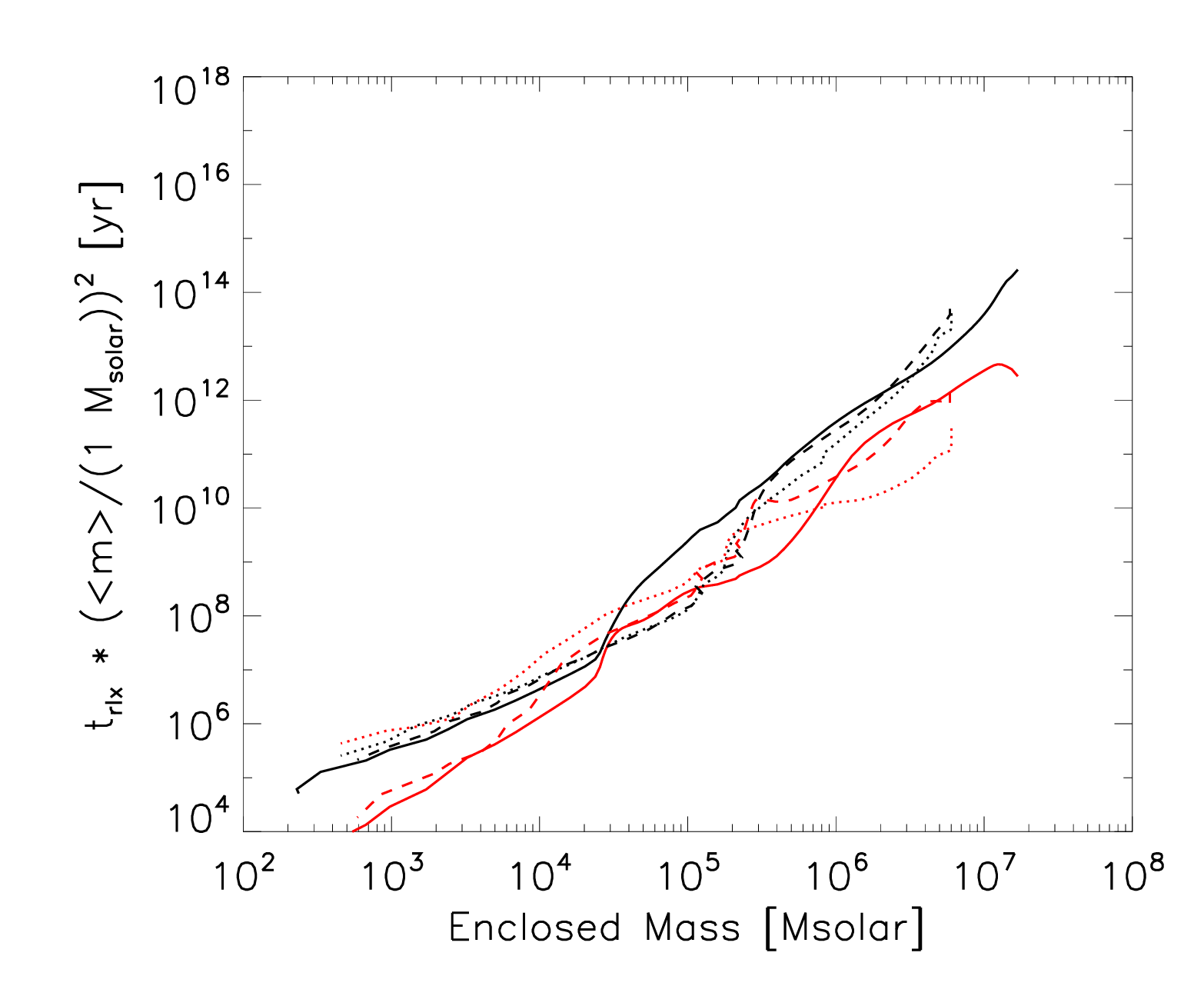}
  \includegraphics[width=8.8cm]{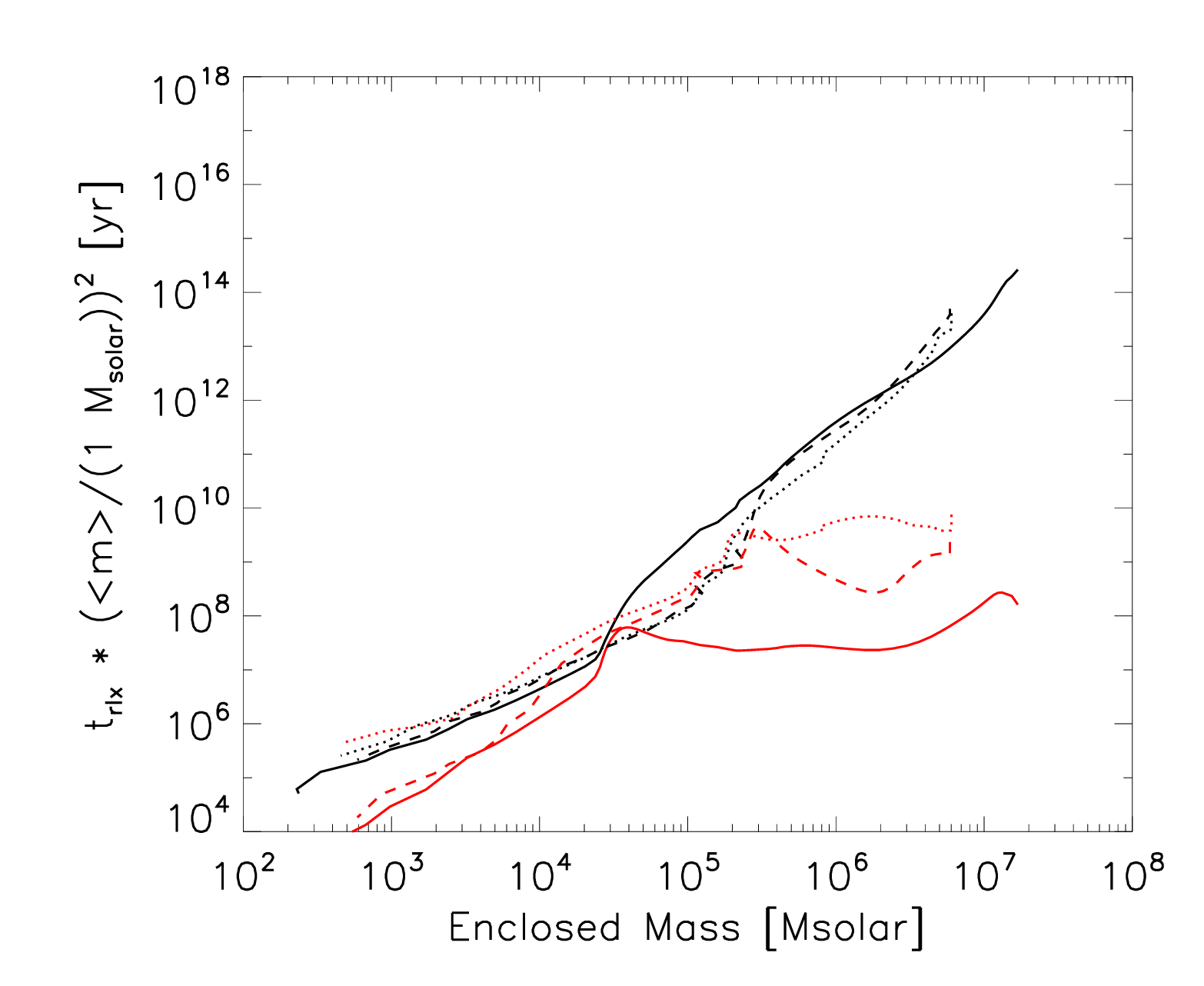}
  \caption[Collision times for the final gas radius in a star
    cluster. ]{\label{Ctimes} {\it Top Panel:} {\it Left-hand Panel:} The
    black curves show the expected collision times (equation \ref{tcoll}) for stellar
    collisions in a  compact  star cluster with masses and radii as 
    that of gas at the end of simulation A-C. The red curves are 
    the expected collision times assuming that the gas settles into
    rotational support without losing further angular momentum 
    using the values of $R_{\rm{final}}$ from the left hand panel of Figure
    \ref{Rfinal}.  {\it Right-hand Panel:} The collision times assuming
    that further angular momentum is lost as described in section 3.3 
    and using values of $R_{\rm{final}}$ as  in the right hand panel of Figure \ref{Rfinal}.
    {\it Bottom Panel:} The same plots except in this case the relaxation
    time (equation \ref{trlx})  is plotted against enclosed mass. 
}

\end{figure*}
 

\subsection{Super-massive stars and quasi-stars}

At the time when the gas settles into rotational support in our
simulation the surface mass density at the center exceeds $30 \, \rm{g/cm^2}$ 
(Figure 1) and the gas is thus  optically thick. The dominant energy production
process is the release of gravitational energy. As argued in \S \ref{timescales} 
this will occur on a few times the dynamical time scale. The
corresponding mass inflow  rate and luminosity 
rise rapidly as the gas contracts. They  can be estimated as 
$\dot M \sim 0.1 \, v_{\rm c}^{3}/G
\sim 1 \, (v_c/30 \kms)^3 \, M_{\odot} \, \rm{yr}^{-1}$ and 
$L_{\rm grav} \sim 0.1 \, v_{\rm c}^{5}/G \sim 3.6 \, (v_c/30 \kms)^5 \times 10^{38}
\, \rm{erg \, s^{-1}}$ . If the gas in the discs in our simulations is able to
contract by another factor 100 in radius without
efficient fragmentation the circular velocity will rise to $300\kms$.
At this stage the luminosity due to the release of gravitational
energy will become comparable  to the Eddington luminosity $L_{\rm Edd}$,
where the outward pressure  due to the radiation 
balances the  gravitational attraction,
\begin{equation}
L_{\rm Edd} = {4 \pi G M m_H c \over \sigma_T} \simeq 1.3 \times 10^{43} 
( {M \over 10^5 M_{\odot}} ) \, \rm{erg \, s^{-1}}  , 
\end{equation}

where c is the speed of light, $m_H$ is the mass of a hydrogen atom  and $\sigma_T$ 
is the Thomson scattering cross-section.

\indent If the entropy produced due to the release of gravitational binding energy 
can be trapped in a quasi-spherical
configuration  the disc should puff up \cite[e.g.][]{Begelman_1978} 
when the luminosity due to the release of gravitational binding energy
exceeds the Eddington luminosity and a (radiation)
pressure supported super-massive 
star should  ensue.  \cite{Begelman_2006,Begelman_2008} 
have argued that a supermassive star embedded in a spherical
accretion flow with mass accretion rates in the range 
$\sim 0.1 -1  \, M_{\odot} \, \rm{yr^{-1}}$ will develop a core-envelope 
structure akin to that of a Thorne-Zytkow object where a compact
neutron star is embedded in a diffuse red giant star \cite[]{Thorne_1977}. \\
\indent \cite{Begelman_2006} dubbed such an object a  ``quasi-star''
and envisage its evolution as follows.
Initially a compact core of about $10 - 20 \, M_{\odot}$ surrounded by a more
diffuse  radiation-pressure  supported envelope forms. 
In the central regions nuclear reactions  may
commence but the corresponding energy release is not expected to 
exceed  the release of gravitational binding energy.  
The  high mass  in-fall rate continues to compress
and heat the core. \cite{Begelman_2008} argue that the core can reach 
temperatures of up to $5 \times 10^8$ K at which point the gas should cool
catastrophically due to thermal neutrino cooling leading to a
core-collapse black hole. At this stage the black hole is 
surrounded by a massive gas envelope of more than one hundred times
the mass of the black hole  and the
central black hole accretes at a rate which corresponds to the Eddington rate
of the mass in the envelope.  Given that the mass of the envelope 
is much larger than the mass of the black hole 
this allows the  black hole to quickly grow to a mass of $10^3 - 10^4 \,
M_{\odot}$, at which point \cite{Begelman_2008} argue 
the quasi-star  evaporates as the photospheric temperature falls. Whether
this is actually the correct picture is still rather  uncertain but
definitely merits further investigation. It is, however, interesting to note that the 
postulated in-fall rates in the Begelman et al.  model are similar to
those in our  simulations. The
in-fall rates of the gas in the outer mass shells at the end of our
simulations range from $0.3-1.5$ solar masses per year and as we have 
discussed at the start of this section  the inflow rates at the centre
of  the discs, which have formed in our simulations, are expected to
become even larger as the self-gravitating discs evolve on a viscous
timescale due to gravitational instabilities.

\subsection{Super-Eddington accretion of stellar mass black hole
seeds} \label{Eddgrowth}

If some star formation occurs in
the discs at the centre of our haloes a stellar mass black hole 
with a mass of up to 100 $M_{\odot}$ should easily form.
Alternatively a stellar mass black hole which has formed in the early
stages of the collapse or in one of the precursors of our DM haloes may sink to
the centre of the DM potential in which our discs form. Independent of whether a 
supermassive star will form from the gas settling to the centre of the disc  
such black holes find themselves in an ideal environment for rapid growth. 
The surrounding material is flowing to the centre 
at highly super-Eddington accretion rates  compared to the
Eddington accretion of a stellar mass black hole. The rotational velocities
of our discs range from $30-60 \, \kms$ and the material is thus already
strongly bound.   The energy released by a stellar mass black hole will
not be able to  halt the surrounding inflow of matter and the black hole should
grow quickly at highly super-Eddington rates to masses comparable
to the disc masses if fragmentation is not important.  
In Figure \ref{Vcfinal} we plot the rotational velocities at which the outer mass 
shells are expected to settle into rotational support for our two
assumptions regarding angular momentum loss. If the gas in the outer
mass shells lose angular momentum as efficiently as the gas in the discs in
our simulations the rotational velocity could be as large as several
hundred $\kms$. Thi is  quite
remarkable considering that the virial velocities of the DM haloes are
in the range $20-30 \, \kms$.  There is thus potential for rapid further 
growth and the final black hole mass would probably be set by
competition between gas consumption by star formation and accretion for the outer 
mass shells or by the increasing feedback due to the increasing
accretion luminosity. \\
\indent 
If the entropy produced due to the release of gravitational energy 
during the accretion is trapped in a  quasi-spherical configuration
then the  accreting  black hole would actually be very similar to the 
later stages of a supermassive star discussed in the last section. 
The released gravitational energy may, however,  not efficiently couple 
to the surrounding inflowing matter. It could instead be transported
outwards  in the form of a jet as is observed in young stars, active
galactic nuclei  and 
probably also gamma ray bursts \cite[e.g.][]{Mundt_1990, Begelman_1984, Meszaros_2002}. 
Accretion would then occur in a fat accretion disc at
super-Eddington accretion rates possibly    
involving the photon bubble instability \cite[]{Gammie_1998,
 Turner_2005} and/or advection  dominated in-/outflows 
as discussed by \cite{Blandford_1999} and \cite{Begelman_2001}.

\begin{figure*}

  \includegraphics[width=8.8cm]{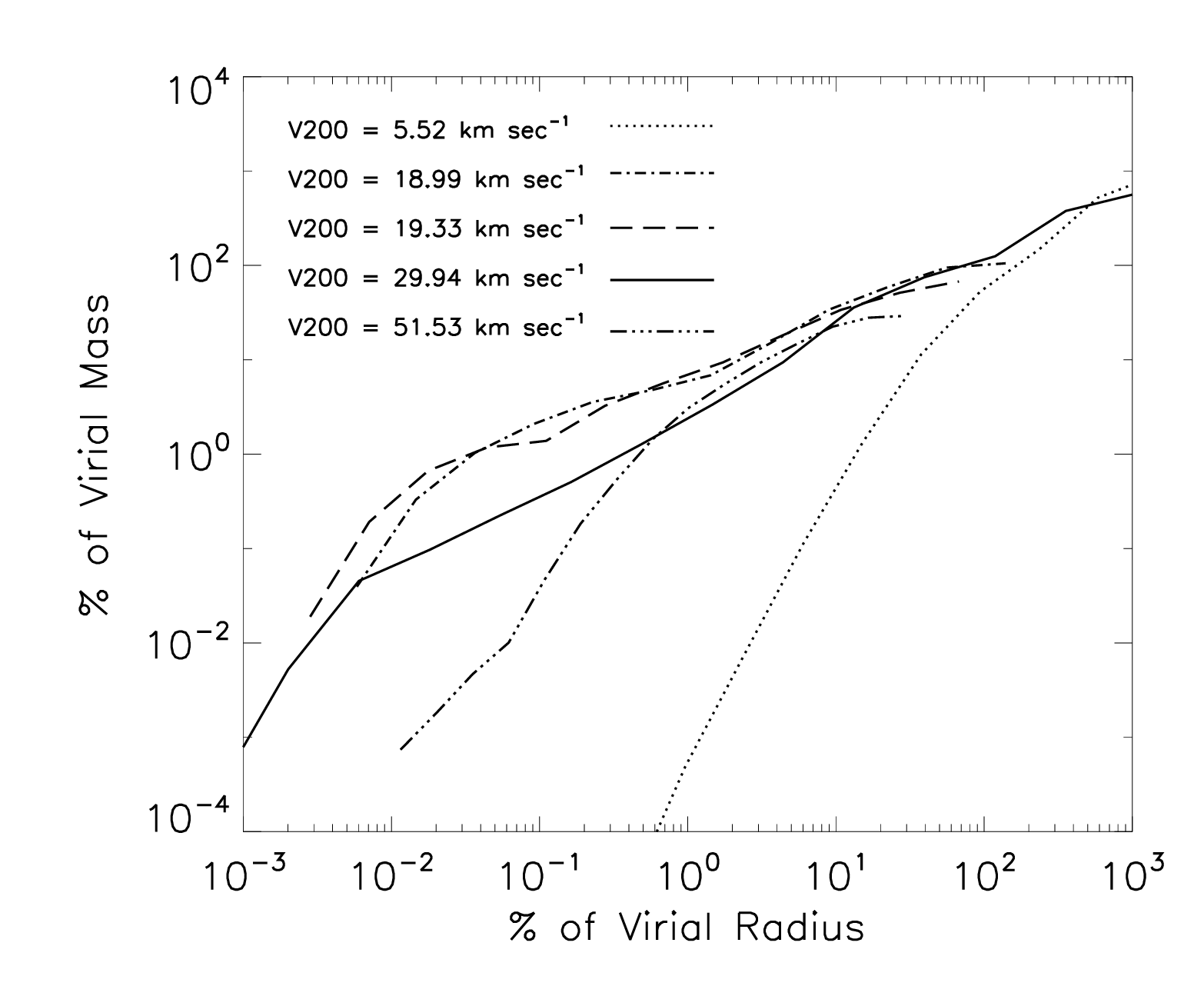}
  \includegraphics[width=8.8cm]{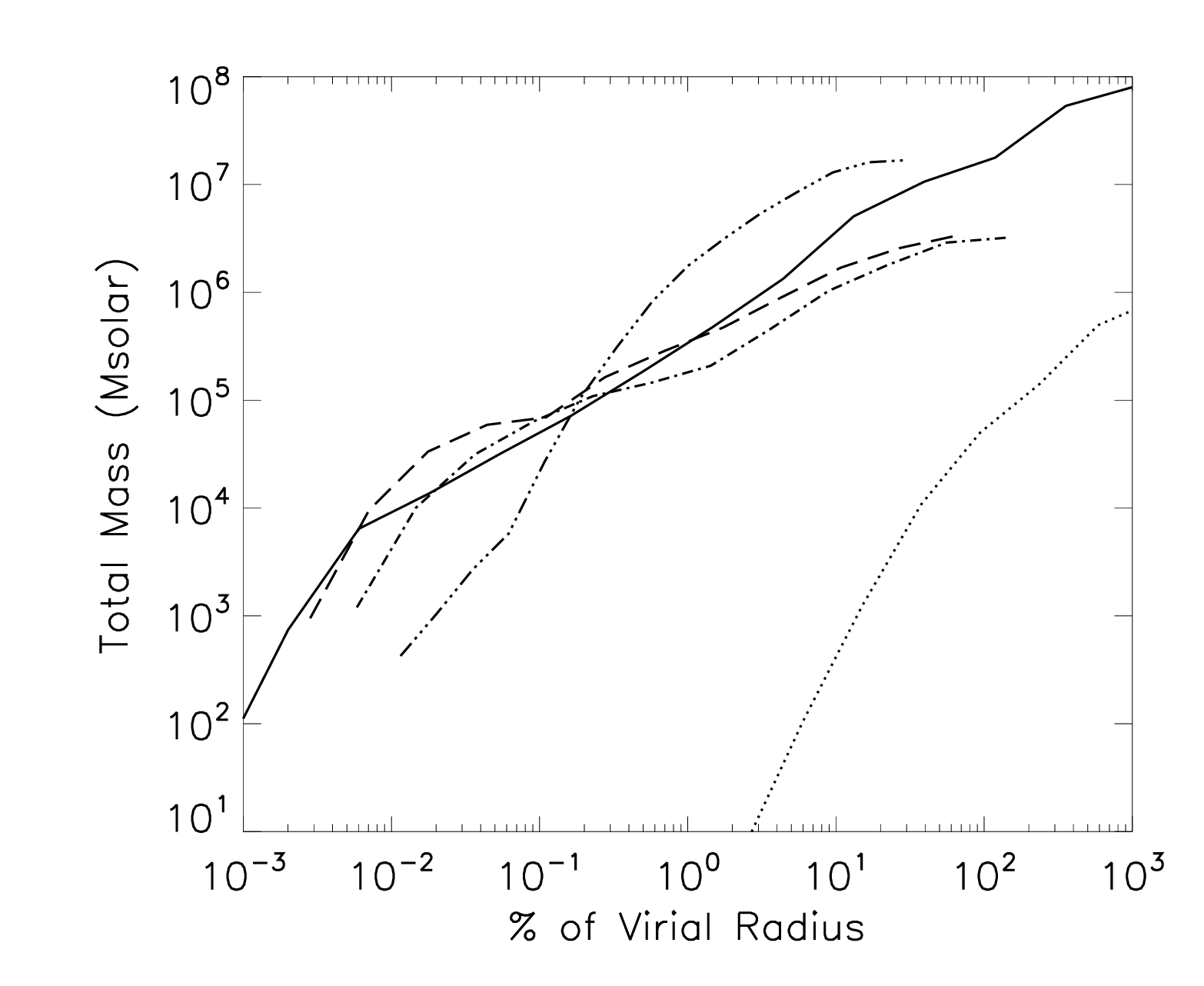}
  \caption[Mass fraction versus fraction of virial
    radius]{\label{Massfrac} {\it Left-hand Panel}: The fraction of
    enclosed mass as a function  of the fraction of the  virial
    radius. 
    {\it Right-hand Panel}: The enclosed  mass as a function of the
    fraction  of the virial radius. The collapse into a compact
    configuration appears to be most efficient in haloes with 
    $V_{\rm vir} \sim 20 \kms$ suggesting that massive seed black holes 
    form most efficiently in these haloes. This should imprint 
    a characteristic mass  on the mass spectrum of massive seed black
    holes.}
\end{figure*}


\subsection{The formation and evolution of a compact star
cluster} \label{cluster}

As briefly discussed in \S \ref{H2cooling} the evolution of the gas in
dynamical haloes with virial temperatures  $T_{\rm{vir}} \gtrsim 10000 \,
\rm{K}$ depends crucially on whether metals and/or $H_2$ are present
in sufficient  abundance to facilitate cooling well below the virial
temperature. If this is the case the gas at least  in principle can
fragment and form stars, even though it is  unclear how efficient
fragmentation  and star formation in this situation would be \cite[e.g.][]{Omukai_2008}. If
star formation occurs early on in the collapse  the outcome is
probably similar to nearby dwarf galaxies, which have DM haloes of
similar virial velocity  and temperature as those we have found in our
simulations. Some of the nearby dwarf galaxies  should have actually formed  at
similarly high redshifts ($z \sim 15$) as we are discussing here.
Early fragmentation requires the presence of a substantial
metallicity and most likely also dust. Dwarf galaxies have typical
metallicities of a hundredth to a tenth of solar.  Star formation
appears to be very inefficient in dwarf galaxies. This is
generally assumed  to be due to feedback from supernovae and or
ionizing radiation which appears to strongly affect  star formation in
haloes with $V_{\rm c} \lesssim 50 \kms$. \\   
\indent It seems, however,
likely that at $z\sim 15$ (and perhaps even at the present day) 
a fraction of DM haloes with virial temperatures $T_{\rm
vir} \gtrsim 10000 \, \rm{K}$  have not yet been significantly
polluted with metals or that the metals in them have not been sufficiently mixed into
the cold gas phase for metal cooling  to be relevant. \\
\indent In such haloes stars could probably only form by $H_2$
cooling once rotational  support has set in.  As discussed in \S
\ref{pathways}  this would lead to very compact star clusters  which
may well be substantially more compact (see Figure \ref{Rfinal}) than
the globular clusters and the compact nuclear  star clusters  which have
been discovered in the centres of many  galaxies including luminous elliptical galaxies and
the bulges of spiral galaxies thanks to the unrivaled resolution of the Hubble
Space Telescope  \cite[e.g.][]{Carollo_1997,
Carollo_1998, Boker_2002, Cote_2006}. \\ 
\indent If the gas in our
haloes were to fragment and form a star cluster, after rotational support
sets in,  then the typical
velocity dispersions of the star cluster would be substantially
larger than the  velocity dispersion of the DM haloes which host them
and could in  principle reach several hundred $\kms$ if the outer mass
shells lose angular momentum as  efficiently as the gas in the discs
in our simulations.\\  
\indent Runaway collision at the the centre of
dense stellar systems is another promising  pathway to the formation of a
massive black hole which has been studied intensively (Begelman \&
Rees 1978; \nocite{Begelman_1978} see  \cite{Freitag_2008} for a
review and further references).  For a stellar cluster with density
$n_*$ and velocity dispersion  $\sigma_{\rm rel}^2 = \sigma_1^2+\sigma_2^2$
with two populations of stars with stellar masses $m_1$ and $m_2$ and
stellar radii $r_1$ and $r_2$ the collision time can be estimated as, 
\cite[]{Binney_1987, Freitag_2008}
\begin{equation}  \label{tcoll}
t_{\rm coll} \simeq 5 \, \rm{Gyr} \, {1 \times 10^6 \,
\rm{pc^{-3}} \over n_*} {\sigma_{\rm rel} \over 10 \kms} {2 \,
R_{\odot} \over r_1 + r_2} {2 \, M_{\odot} \over m_1 + m_2}.
\end{equation} In Figure \ref{Ctimes} we show this estimate for the
collision time scale of putative star clusters forming when the gas
would settle into rotational support  in our simulations  for our two
assumptions regarding the loss of angular momentum. 
We  take $3 M/4\pi R^{3}$ as a proxy for the stellar mass density
and estimate $\sigma_{\rm rel}$ as  $V_c/\sqrt{2}$.  For solar mass stars the
collision timescales for a star cluster, with mass and radius
similar to that expected for the gas  in our simulations if it 
continues to loose angular momentum,  could become  shorter than  
the Hubble time  ($\sim 4.16 \times 10^8 \,
\rm{yr}$ at $z = 15$). \\
\indent Stars in a populous cluster ($N \gg 10$) 
exchange  energy and angular momentum 
through two-body relaxation. The result of this  relaxation
process is the efficient mass segregation of the cluster with more
massive stars concentrated at the centre of the cluster
and a much enhanced central density 
\cite[e.g.][]{Spitzer_1969, Binney_1987}
which eventually can lead to core collapse.
The estimate in   Figure \ref{Ctimes}  
will thus be  a substantial overestimate for the 
collision time of the central regions of such a 
cluster especially if core collapse has occurred.\\
\indent The relaxation time can  be estimated as in \cite{Freitag_2008}, 
\begin{equation} \label{trlx}
t_{\rm{rlx}} = 2 \, \rm{Myr} \, \frac{10}{\Lambda}
\frac{10^6 pc^{-3}}{n_*} \Big({\sigma_{\rm rel} \over 10 \kms }\Big)^3
\Big({1 \, M_{\odot} \over <m>}\Big)^2 
\end{equation} where $\Lambda \simeq ln(0.02 \, N)$, $N$ is the number
of stars in the cluster and $<m>$ is the mean stellar mass.
The timescale for core collapse to occur is a few times the relaxation
timescale at the centre of the cluster. 
If the timescale for core collapse is shorter than the 
main sequence life time of massive stars a very
massive star should form by collisional runaway collapse \cite[]{Gurkan_2006, PortegiesZwart_2004}. \\
\indent In the bottom panel of Figure \ref{Ctimes} we have plotted the
relaxation time for our two assumptions of angular momentum loss. 
The central relaxation time scale and thus the time scale for core collapse 
of a putative cluster  will again be substantially smaller than this. 
Conditions for the formation of a very massive star by collisional
runaway and a subsequent collapse to a  massive black hole appear thus
to be favourable especially for stellar clusters in the lower mass range around
$10^{4}-10^{5} M_{\odot}$\cite[]{Freitag_2008}. 
As discussed in \S \ref{Eddgrowth} such a
massive  black hole would find itself  in an ideal  environment for
rapid further growth.

\subsection{Characteristic masses  of massive black hole seeds} \label{characteristicmasses}

So far we have concentrated on simulations A-C for DM haloes with virial velocities in 
the range $19-30 \, \kms$. Recall that the mass of the discs in our simulations are $2\times 10^4 \, M_{\odot}$ 
and $1\times 10^5 \, M_{\odot}$. Intriguingly the more massive discs form in the smaller mass 
haloes. In order to investigate if this is a general trend we have run  two  further simulations,
D and E, for haloes with larger and smaller masses (see table \ref{TableSims} for details) extending 
the range of virial masses from $\sim 1 \times 10^6 - 1 \times 10^9 \, M_{\odot}$ which correspond to 
virial velocities $\sim 6 - 52 \, \kms$. Due to the  adaptive nature of the Enzo code the spatial
resolution and  the gas resolution is comparable in all simulations.\\
\indent The fraction of the mass within a certain fraction of the virial radius 
at the time when rotational support sets is shown  in the left hand
panel of Figure \ref{Massfrac}. It  appears to have  a maximum 
for virial velocities around $20 \kms$.
 
For lower mass haloes the temperatures attainable by atomic cooling 
is above the virial temperature of the halo and the gas cannot
collapse without efficient metal or molecular hydrogen cooling.  
Surprisingly for the  more massive haloes the collapse also becomes 
less efficient.  \\
\indent In the right hand panel of Figure \ref{Massfrac} we have plotted the total mass
against  the fraction of  the virial radius. The 
collapse into a compact configuration appears indeed to be most efficient 
in DM haloes with  virial velocities around $20 \kms$ and to lead to the 
formation of compact self-gravitating discs  with characteristic masses of  
$10^5 \, M_{\odot}$. This should result in  a peak in the mass function 
of black hole seeds and establish a characteristic mass for the
population of massive black hole seeds.  Predicting the space density of massive black
hole  seeds formed in the way discussed here would require to predict
the fraction of haloes with $V_{\rm vir} \sim 20 \kms$ for which the
conditions for  the suppression of $H_2$ and metal cooling are
favourable which is beyond the scope of this paper. However a 
fraction as small as 0.1\% would be sufficient to seed the 
population of present-day super-massive black holes in galactic
bulges.

\section{Discussion and Conclusions} \label{conclusions}

We have discussed here the pathways to massive black holes in 
DM haloes with $T_{\rm vir} \gtrsim 10000 \, \rm{K}$ based on insights
into the formation of self-gravitating compact discs from \enzo  
AMR simulations of  atomic cooling driven  collapse in 
such haloes.  \\
\indent We argue that in a  fraction of these  haloes the conditions
should be just right for a rapid and  efficient build-up of 
massive black holes and  rather different from those found in the 
DM  host haloes of dwarf galaxies with similar virial velocities. \\
\indent For this to be the case the metallicity in the collapsing gas in these haloes has to 
be sufficiently low for  metal cooling not to induce early efficient 
fragmentation and star formation.  The gas can then lose 90\% and more of its angular
momentum due to supersonic turbulent motions before it first settles into 
rotational support in a very compact configuration with rotational
velocities well in excess of the  virial velocity of the DM halo. 
The corresponding fat, self-gravitating discs are marginally stable 
against gravitational  instabilities. \\
\indent If at this stage either $H_2$ or metal cooling induce sufficient 
fragmentation/star formation a very  compact nuclear star 
cluster  would form offering favourable conditions for the 
formation of an intermediate mass black hole by collisional run-away
possibly with the intermediate stage of a supermassive star. \\
\indent If instead after the onset of rotational support  atomic cooling is still the
dominant cooling process because the metallicity is still low and $H_2$
cooling is suppressed  due to external or more likely
internal UV Lyman-Werner radiation the gas will still be unable
to fragment while  gravitational instabilities will induce 
continued  loss of  angular momentum. \\
\indent If the entropy due to the released gravitational energy can thereby
be efficiently trapped the geometry of the inflow of gas  will stay  
spherical and a supermassive star will form. The ongoing
collapse of the outer mass shells occurs with mass accretion rates of a 
few tenth to a few $M_{\odot} \, \rm{yr^{-1}}$.  This is  just the range
of accretion rate for which \cite{Begelman_2006, Begelman_2008} have suggested that the 
supermassive star should take the form of a rapidly accreting
quasi-star with a pronounced core-envelope structure and a rapidly 
growing core which will turn into a rapidly growing stellar mass black
hole. \\
\indent If the gravitational binding energy is instead transported away 
in a disc-jet configuration the accretion flow driven by the in-fall of 
the outer mass shells is likely to be directed onto a remnant black hole 
of an ordinary star of which the DM halo should contain at least a
few.\\
\indent In any case the infalling gas becomes strongly bound and has inflow
rates which largely exceed  the Eddington accretion rate for a stellar
mass black hole. Feedback due to individual supernova is expected
to have little direct effect on the dynamical  evolution of the gas at 
this stage and  rapid growth to a massive black hole is expected. \\
\indent We have performed numerical simulations for a range of virial
velocities from $6$ to  $52 \, \kms$. As expected in DM haloes with  
$V_{\rm{vir}} \lesssim 10 \kms$ the temperature of the gas attained by
atomic cooling only is too low for
the gas to collapse. Somewhat surprisingly in  the more massive haloes
with masses above this threshold the settling  of the gas at the
centre of the DM haloes becomes rapidly less efficient with increasing 
mass. If the  gas  temperature 
is well below  the virial temperature of the DM halo the gas appears to break more
easily into  several lumps and random motions become more
efficient  in stabilizing against collapse.   
DM haloes with virial velocities close to the threshold of 
$V_{c} \gtrsim 10 \kms$ show the largest gas mass settling
into a self-gravitating disc. \\
\indent This suggests that massive seed
black holes will form with a characteristic mass. The exact value
of this  characteristic mass is uncertain but should be set either
by the competition  of accretion onto the massive black hole and
consumption of gas in star formation or by the feedback on the inflow
from the energy released in the accretion flow;
$10^{5}-10^{6} \, M_{\odot} $  appears to be a likely mass range. 
The population of massive seed black holes forming in pre-galactic haloes 
at  $z\sim 15$ discussed here will  naturally grow into
the observed population in observed galactic bulges during the rapid 
hierarchical build-up of galaxies expected to occur at somewhat
smaller redshift. The early formation of such massive 
black holes  strongly alleviates the many problems which have been identified 
if supermassive black holes had to grow from stellar mass black holes 
in smaller dark matter haloes or more generally in an environment where 
supernova feedback from  star formation can prevent the supply of 
sufficient amounts of  fuel for rapid growth.    

\section*{Acknowledgements}
We thank Mitch Begelman, Giuseppe Lodato, Prija Natarajan, Jim
Pringle, Martin Rees and Chris Tout for helpful discussions. 
The numerical simulations were performed on the \cosmos (SGI Altix 3700) supercomputer
at DAMTP in Cambridge and on the Cambridge High Performance Computing 
Cluster \darwin in Cambridge. \cosmos is a UK-CCC facility
which is supported  by HEFCE and PPARC. \darwin is the primary supercomputer 
of the University of Cambridge High Performance Computing Service 
(http://www.hpc.cam.ac.uk/), provided by Dell Inc. using Strategic 
Research Infrastructure Funding from the Higher Education Funding 
Council for England. Special thanks to Amanda Smith for using her artistic wizardry in 
preparing Figure \ref{cartoon}.

\bibliographystyle{mn2e}

\end{document}